\documentclass[fleqn,usenatbib]{mnras}

\usepackage{newtxtext,newtxmath}

\usepackage[T1]{fontenc}
\usepackage{ae,aecompl}

\usepackage{graphicx}	
\usepackage{amsmath}	
\usepackage{amssymb}	
\usepackage{bm}       
\usepackage{etoolbox}
\makeatletter 
  \patchcmd{\NAT@citex}
    {\@citea\NAT@hyper@{%
      \NAT@nmfmt{\NAT@nm}%
      \hyper@natlinkbreak{\NAT@aysep\NAT@spacechar}{\@citeb\@extra@b@citeb}%
      \NAT@date}}
    {\@citea\NAT@nmfmt{\NAT@nm}%
    \NAT@aysep\NAT@spacechar\NAT@hyper@{\NAT@date}}{}{}

  \patchcmd{\NAT@citex}
    {\@citea\NAT@hyper@{%
      \NAT@nmfmt{\NAT@nm}%
      \hyper@natlinkbreak{\NAT@spacechar\NAT@@open\if*#1*\else#1\NAT@spacechar\fi}%
        {\@citeb\@extra@b@citeb}%
      \NAT@date}}
    {\@citea\NAT@nmfmt{\NAT@nm}%
    \NAT@spacechar\NAT@@open\if*#1*\else#1\NAT@spacechar\fi\NAT@hyper@{\NAT@date}}
    {}{}
\makeatother

\newcommand\colt{\textsc{colt}} 
\newcommand\HI{{H\,\textsc{i}}} 

\title[Discrete diffusion Ly$\alpha$ radiative transfer]{Discrete diffusion Lyman $\balpha$ radiative transfer}

\author[A.\ Smith et al.]{
  Aaron~Smith,\thanks{E-mail: \href{mailto:asmith@astro.as.utexas.edu}{asmith@astro.as.utexas.edu}}
  Benny~T.-H.~Tsang,
  Volker~Bromm and
  Milo\v{s} Milosavljevi\'{c}
  \\
  Department of Astronomy, The University of Texas at Austin, Austin, TX 78712, USA
}

\date{Accepted XXX. Received YYY; in original form ZZZ}

\pubyear{2017}

\begin{document}
\label{firstpage}
\pagerange{\pageref{firstpage}--\pageref{lastpage}}
\maketitle

\begin{abstract}
  Due to its accuracy and generality, Monte Carlo radiative transfer (MCRT) has emerged as the prevalent method for Ly$\alpha$ radiative transfer in arbitrary geometries. The standard MCRT encounters a significant efficiency barrier in the high optical depth, diffusion regime. Multiple acceleration schemes have been developed to improve the efficiency of MCRT but the noise from photon packet discretization remains a challenge. The discrete diffusion Monte Carlo (DDMC) scheme has been successfully applied in state-of-the-art radiation hydrodynamics (RHD) simulations. Still, the established framework is not optimal for resonant line transfer. Inspired by the DDMC paradigm, we present a novel extension to \emph{resonant} DDMC (rDDMC) in which diffusion in space and frequency are treated on equal footing. We explore the robustness of our new method and demonstrate a level of performance that justifies incorporating the method into existing Ly$\alpha$ codes. We present computational speedups of $\sim 10^2$--$10^6$ relative to contemporary MCRT implementations with schemes that skip scattering in the core of the line profile. This is because the rDDMC runtime scales with the spatial and frequency resolution rather than the number of scatterings---the latter is typically $\propto \tau_0$ for static media, or $\propto (a \tau_0)^{2/3}$ with core-skipping. We anticipate new frontiers in which on-the-fly Ly$\alpha$ radiative transfer calculations are feasible in 3D RHD. More generally, rDDMC is transferable to any computationally demanding problem amenable to a Fokker-Planck approximation of frequency redistribution.
\end{abstract}




\section{Introduction}
\label{sec:intro}
The Lyman $\alpha$ (Ly$\alpha$) line of neutral hydrogen~(\HI) is an important probe of galaxy formation and evolution throughout cosmic history \citep{Partridge_Peebles_1967}. However, due to the complex nature of resonant scattering of Ly$\alpha$ photons in optically-thick environments, the necessary radiative transfer modeling and interpretation of observations are often challenging \citep{Dijkstra_2014}. Encouragingly, the fundamental physical processes are well studied and a few analytic solutions exist in the literature for idealized cases \citep{Harrington_1973,Neufeld_1991,Loeb_Rybicki_1999,Dijkstra_2006,Tasitsiomi_2006b,Higgins_2012}. Furthermore, the development of Monte Carlo radiative transfer (MCRT) codes with acceleration schemes has allowed for an accurate, universal approach to Ly$\alpha$ calculations \citep[e.g.][]{Auer_1968,Ahn_2002,Zheng_2002}. The MCRT method is well-suited for analyzing 3D hydrodynamical simulations in post-processing \citep[e.g.][]{Tasitsiomi_2006,Laursen_2009,Verhamme_2012,Smith_2015}, the exploration of parameter space in empirical modeling \citep[e.g.][]{Gronke_2015}, and spherically-symmetric radiation hydrodynamics \citep[RHD;][]{Smith_CR7_2016,Smith_RHD_2017}. Still, increasingly efficient use of computational resources will undoubtedly extend the utility and scope of Ly$\alpha$ modeling to include more dimensions, parameters, and photon packets for better resolution and more robust convergence.

Historically, discrete-ordinate and grid-based methods were employed to calculate Ly$\alpha$ radiative transfer solutions for 1D geometries. Eventually, the Monte Carlo approach emerged as the prevalent method due to its accuracy and generality, although several studies continue to explore alternatives by numerically solving approximations of the Ly$\alpha$ radiative transfer equation \citep[e.g.][]{Roy_2010,Yang_2011,Yang_2013,Higgins_2012}. A drawback of these and other techniques based on directional and frequency discretization is the difficulty of adequately representing frequency redistribution, especially when accounting for angular dependence. In many cases these methods are also restricted to slab or spherical symmetry, velocity monotonicity, time-independence, artificial boundary conditions, or require grids and interpolation schemes tailored to specific applications. MCRT codes overcome this by employing continuous, stochastic sampling to build physical quantities from numerous statistical realizations, adapting the accuracy to the available computational resources.

Most Ly$\alpha$ MCRT codes employ a core-skipping technique to accelerate frequency diffusion into the wings of the Ly$\alpha$ profile, thereby facilitating spatial excursion albeit indirectly. These schemes work in optically-thick regions where photons near line centre have negligible mean free paths, and skip to a scattering with a rapidly moving atom that brings the photon out of the core \citep{Auer_1968,Avery_1968,Ahn_2002}. However, \citet{Tasitsiomi_2006} and \citet{Laursen_2009} further incorporated a semi-analytic scheme based on an approximate Neufeld solution to reassign frequencies and directions of photons when the optical depth based on the distance to the nearest face of a resolution element is extremely large. In practice such random walk prescriptions and their variants lose efficiency in simulations with high spatial resolution due to close proximity to cell interfaces where the acceleration scheme is not active \citep{Gentile_2001}. Furthermore, the robustness of modified walks is not well understood in the case of line transfer. For example, the threshold criterion for triggering the acceleration scheme only considers the line centre opacity ($a \tau_0 \gg 1$), which is problematic for wing photons as they do not retain directionality under consecutive applications of the random walk acceleration scheme. In this paper we employ a different method for direct spatial and frequency diffusion to significantly accelerate Ly$\alpha$ MCRT, as described below.

Discrete Diffusion Monte Carlo (DDMC) techniques have been developed to increase the efficiency of MCRT calculations in opaque media \citep{Gentile_2001,Densmore_2007}. The idea is to replace many unresolved Monte Carlo scatterings with a single jump to a neighboring cell based on a discretized diffusion equation. Under the Fick's law closure relation of the radiative transfer equation, the diffusive term operates as spatial leakage across cell interfaces that may be naturally incorporated into the Monte Carlo paradigm. In practical applications, DDMC is only used if the cell optical depth is above a threshold value, e.g. $\tau_\nu \gtrsim 5$. DDMC packets are converted to normal Monte Carlo packets when the frequency-dependent cell optical depth drops below the threshold due to a change of the absorption coefficient along the packet's trajectory. The DDMC method was originally developed for monochromatic thermal radiative transfer. It has recently been extended to incorporate frequency-dependent transfer \citep{Abdikamalov_2012,Densmore_2012,Wollaeger_2013,Wollaeger_2014}. The method has also been applied to the study of neutrino transport in supernovae \citep{Wollaeger_2017} and radiative feedback in massive star cluster formation \citep{Tsang_2017}.

Diffusion acceleration schemes may be necessary for 3D RHD simulations to resolve the essential physical scales on which the characteristics of Ly$\alpha$ transport are set. In fact, the computational efficiency of Ly$\alpha$ MCRT is especially relevant in RHD contexts incorporating resonant scattering in highly time-variable, optically thick environments. Recently, \citet{Smith_DCBH_2017} made a case for the dynamical importance of Ly$\alpha$ radiation pressure in sites where very massive stars formed from primordial gas collapse directly into black holes \citep[reviewed in][]{Volonteri_2012,Smith_AG_2017}. They carried out an exploratory post-processing analysis of a high-resolution ab initio cosmological simulation, concluding that fully coupled Ly$\alpha$ RHD will be crucial to consider in the future because Ly$\alpha$ photons provide a significant source of dynamical feedback. There are also other astrophysical settings where Ly$\alpha$ RHD effects may be important, including Population~III stars \citep[e.g][]{McKee_Tan_2008,Stacy_2012,Stacy_2016}, early quasar formation \citep[e.g.][]{Haehnelt_1995,Milosavljevic_2009}, and interstellar/galactic winds \citep[e.g.][]{Dijkstra_Loeb_2008,Smith_RHD_2017}. The DDMC method can also be utilized to accelerate Ly$\alpha$ MCRT transfer even when the radiation pressure is negligible, both in addition to core-skipping and in cases where core-skipping is not desired. Therefore, it seems timely to apply the DDMC method to Ly$\alpha$ radiative transfer problems, to which we devote the remainder of this paper.

The paper is organized as follows.  In Section~\ref{sec:methods}, we provide the background for Ly$\alpha$ transfer and describe the DDMC methodology within this context. In Section~\ref{sec:tests}, we test the method against analytical solutions and explore the time-dependent behavior in homogeneous media. In Section~\ref{sec:performance}, we discuss the robustness of the DDMC framework and demonstrate strong performance characteristics compared to traditional MCRT. In Section~\ref{sec:developments}, we comment on several aspects of combined discrete and continuous transport schemes. In Section~\ref{sec:conc}, we explore potential applications of the DDMC method to future Ly$\alpha$ studies.

\section{Methodology}
\label{sec:methods}
In this section we summarize the fundamental concepts behind the DDMC method and the specific details relating to its application to Ly$\alpha$ radiative transfer.

\subsection{Lyman-\texorpdfstring{$\balpha$}{α} radiative transport}
The specific intensity $I_\nu(\bmath{r}, \bmath{n}, t)$ encodes all information about the radiation field taking into account the frequency $\nu$, spatial position $\bmath{r}$, propagation direction unit vector $\bmath{n}$, and time $t$. The general Ly$\alpha$ radiative transfer equation is given by
\begin{equation} \label{eq:RTE}
  \frac{1}{c} \frac{\partial I_\nu}{\partial t} + \bmath{n} \bmath{\cdot} \bmath{\nabla} I_\nu = j_\nu - k_\nu I_\nu + \iint k_{\nu'} I_{\nu'} R_{\nu', \bmath{n}' \rightarrow \nu, \bmath{n}} \text{d}\Omega' \text{d}\nu' \, ,
\end{equation}
where $k_\nu$ is the absorption coefficient, $j_\nu$ is the emission coefficient, and the last term accounts for frequency redistribution after partially coherent scattering \citep{Dijkstra_2014}. The redistribution function $R$ is the differential probability per unit initial photon frequency $\nu'$ and per unit initial directional solid angle $\Omega'$ that the scattering of such a photon traveling in direction $\bmath{n}'$ would place the scattered photon at frequency $\nu$ and directional unit vector $\bmath{n}$ \citep[for the historical development see][]{Henyey_1940,Zanstra_1949,Unno_1952}.

It is convenient to convert to the dimensionless frequency $x \equiv (\nu - \nu_0) / \Delta \nu_\text{D}$, where $\nu_0 = 2.466 \times 10^{15}$~Hz is the frequency of the Ly$\alpha$ transition, $\Delta \nu_\text{D} \equiv (v_\text{th}/c) \nu_0$ is the Doppler width of the profile, and the thermal velocity in terms of $T_4 \equiv T/(10^4\,\text{K})$ is $v_\text{th} \equiv (2\,k_\text{B} T / m_\text{H})^{1/2} = 12.85\,T_4^{1/2}\,\text{km}\,\text{s}^{-1}$. Furthermore, the natural Ly$\alpha$ line width is $\Delta \nu_\text{L} = 9.936 \times 10^7$~Hz and the `damping parameter', $a \equiv \Delta \nu_\text{L} / 2 \Delta \nu_\text{D} = 4.702 \times 10^{-4}\,T_4^{-1/2}$, represents the relative broadening of the natural line. This is important because the frequency dependence of the absorption coefficient is encapsulated within the Voigt profile $\phi_\text{Voigt}$. For convenience we define the Hjerting-Voigt function $H(a,x) = \sqrt{\upi} \Delta \nu_\text{D} \phi_\text{Voigt}(\nu)$ as the dimensionless convolution of Lorentzian and Maxwellian distributions,
\begin{equation} \label{eq:H}
  H(a,x) = \frac{a}{\pi} \int_{-\infty}^\infty \frac{e^{-y^2}\text{d}y}{a^2+(y-x)^2} \approx
    \begin{cases}
      e^{-x^2} & \quad \text{`core'} \\
      {\displaystyle \frac{a}{\sqrt{\pi} x^2} } & \quad \text{`wing'}
    \end{cases} \, .
\end{equation}
The approximate frequency marking the crossover from core to wing is denoted by $x_\text{cw}$, i.e. where $\exp(-x_\text{cw}^2) \simeq a / \sqrt{\pi} x_\text{cw}^2$.

Expressions for the redistribution function and discussions of its properties may be found in \citet{Unno_1952}, \citet{Osterbrock_1962}, \citet{Hummer_1962}, and \citet{Lee_1974}. One notational simplification is that the scattering probability depends only on the angle between the incoming and outgoing directions, $\mu \equiv \cos \theta = \bmath{n} \bmath{\cdot} \bmath{n}'$. Therefore, with an appropriate choice of a scattering phase function $p(\mu)$ normalized such that $\int_{-1}^1 p(\mu)\,\text{d}\mu = 1$, we define the outgoing angular-averaged redistribution function as $R_{x' \rightarrow x} \equiv (4\upi)^{-2} \iint \text{d}\Omega' \text{d}\Omega R_{x', \bmath{n}' \rightarrow x, \bmath{n}} = \int_{-1}^1 p(\mu) R_{x' \rightarrow x, \mu}\,\text{d}\mu$. Furthermore, the conservation of photons in equation~(\ref{eq:RTE}) requires a normalization\footnote{This convention differs from that of \citet{Hummer_1962}, but is similar to that of \citet{Dijkstra_2014}. For reference, our angle-averaged definition is related as $R_{x' \rightarrow x}^\text{us} = R_{x' \rightarrow x}^\text{Hummer} / \phi_\text{Voigt}(x')$.} for the redistribution function of $\int_{-\infty}^\infty R_{x' \rightarrow x}\,\text{d}x' = 1$, in accordance with the previous interpretation as a probability distribution function. Finally, the conversion to dimensionless frequency introduces a constant multiplicative factor, e.g. specific intensity, $I_x = \Delta \nu_\text{D} I_\nu$, and frequency redistribution, $R_{x' \rightarrow x} = (\Delta \nu_\text{D})^2 R_{\nu' \rightarrow \nu}$. Further discussion of the behavior of the redistribution function is provided in Appendix~\ref{appendix:redistribution}.

\subsection{Diffusion regime}
We define the zeroth and first order angular moments of the radiation intensity as $J_x \equiv \frac{1}{4\upi} \int \text{d}\Omega I_x$ and $\bmath{H}_x \equiv \frac{1}{4\upi} \int \text{d}\Omega I_x \bmath{n}$. These quantities are related to the energy density and flux by $E_x = \frac{4\upi}{c} J_x$ and $\bmath{F}_x = 4\upi\,\bmath{H}_x$, respectively. The emissivity is discretized as creation of photon MC packets each characterized by a particular energy $\varepsilon_k$, position $\bmath{r}_k$, frequency $x_k$, and emission time $t_k$, such that $j_x \approx \sum \varepsilon_k \delta(\bmath{r}_k) \delta(x_k) \delta(t_k) / (4 \upi)$, where the index $k$ refers to an individual MC packet.  Without loss of generality we omit the emissivity term in the following discussion. The angular-averaged form of equation~(\ref{eq:RTE}) is the zeroth order moment equation:
\begin{equation} \label{eq:RTE-moment}
  \frac{1}{c} \frac{\partial J_x}{\partial t} + \bmath{\nabla} \bmath{\cdot} \bmath{H}_x = -k_x J_x + \int k_{x'} J_{x'} R_{x' \rightarrow x} \text{d}x' \, .
\end{equation}
In the diffusion limit this is further simplified by applying Fick's law as a closure relation to the moment equations:
\begin{equation} \label{eq:Ficks-Law}
  \bmath{H}_x = -\frac{\nabla J_x}{3 k_x} \, ,
\end{equation}
where the factor of $3$ arises from the number of dimensions. Likewise, in optically thick media we can take advantage of the Fokker-Planck approximation to rewrite the redistribution integral as \citep{Rybicki_1994}
\begin{equation} \label{eq:Fokker-Planck}
  -k_x J_x + \int k_{x'} J_{x'} R_{x' \rightarrow x} \text{d}x' \approx \frac{\partial}{\partial x} \left( \frac{k_x}{2} \frac{\partial J_x}{\partial x} \right) \, ,
\end{equation}
which naturally transforms frequency redistribution into a localized diffusion process. We note that the form of equation~(\ref{eq:Fokker-Planck}) technically violates photon conservation, but the correction factor proposed by \citet{Rybicki_1994} with frequency derivatives of $\bmath{n} \bmath{\cdot} \bmath{H}_x$ may safely be ignored for nearly isotropic radiation fields. Furthermore, equation~(\ref{eq:Fokker-Planck}) does not account for detailed balance, atomic recoil, or stimulated scattering. However, these are unlikely to be significant for the applications of this paper \citep{Rybicki_2006}. Thus, after incorporating equations~(\ref{eq:Ficks-Law}) and (\ref{eq:Fokker-Planck}) we have
\begin{equation} \label{eq:RTE-continuous-double-diffusion}
  \frac{1}{c} \frac{\partial J_x}{\partial t} = \nabla \cdot \left( \frac{\nabla J_x}{3 k_x} \right) + \frac{\partial}{\partial x} \left( \frac{k_x}{2} \frac{\partial J_x}{\partial x} \right) \, .
\end{equation}
This form places diffusion in space and frequency on equal footing.

Finite-volume discretization in space and frequency transforms the diffusion terms into source and sink terms dictating the movement of photon packets through cell boundaries and frequency bins. This process is quantitatively described by `leakage coefficients', sometimes referred to as `leakage opacities'. We define the cell- and bin-averaged intensity by $J_{i,j} \equiv(\Delta V_i \Delta x_j)^{-1} \iint_{i,j} J_x\, \text{d}V \text{d}x $, with $\Delta V_i \equiv \int_i \text{d}V$ denoting the volume of cell $i$, and $\Delta x_j \equiv \int_j \text{d}x$ the width of frequency bin $j$. The spatial diffusion term becomes
\begin{equation} \label{eq:discrete-z-term}
  \nabla \cdot \left( \frac{\nabla J_x}{3 k_x} \right) \longrightarrow \sum_{\delta i} k_{z\text{-leak}}^{\delta i} (J_{\delta i,j} - J_{i,j}) \, ,
\end{equation}
where the summation is over all neighboring cells $\delta i$, the cells sharing a face with cell $i$. This discretization of the diffusion operator is based on a continuous piecewise linear reconstruction with inflections at cell centers and interfaces. In the Monte Carlo interpretation the right hand side of equation~(\ref{eq:discrete-z-term}) provides the mechanism for spatial transport. Likewise, the frequency diffusion term is
\begin{equation} \label{eq:discrete-x-term}
  \frac{\partial}{\partial x} \left( \frac{k_x}{2} \frac{\partial J_x}{\partial x} \right) \longrightarrow \sum_{\delta j} k_{x\text{-leak}}^{\delta j} (J_{i,\delta j} - J_{i,j}) \, ,
\end{equation}
where the summation is over neighboring frequency bins $\delta j$.
In the Monte Carlo picture the exchange on the right hand side of equation~(\ref{eq:discrete-x-term}) provides the mechanism for frequency redistribution.

Substituting equations~(\ref{eq:discrete-z-term})~and~(\ref{eq:discrete-x-term}) into equation~(\ref{eq:RTE-continuous-double-diffusion}) yields the fundamental equation for our new DDMC scheme for resonant line transfer with a symmetric treatment of diffusion in both space and frequency:
\begin{equation} \label{eq:RTE-discrete-double-diffusion}
  \frac{1}{c} \frac{\partial J_{i,j}}{\partial t} = \sum_{\delta i} k_{z\text{-leak}}^{\delta i} (J_{\delta i,j} - J_{i,j}) + \sum_{\delta j} k_{x\text{-leak}}^{\delta j} (J_{i,\delta j} - J_{i,j}) \, .
\end{equation}
Equation~(\ref{eq:RTE-discrete-double-diffusion}) is arranged to highlight photon flux conservation across cell and bin interfaces. The relative magnitudes of the leakage coefficients express the likelihood of the respective diffusion events.

The exact form for the leakage coefficients is determined by the specific geometry and discretization scheme. Derivations of $k_{z\text{-leak}}^{\delta i}$ are given by \citet{Densmore_2007} and \citet{Tsang_2017} for non-uniform Cartesian geometry, and \citet{Abdikamalov_2012} for non-uniform spherical geometry. For concreteness, in this paper we employ the following leakage coefficients for non-uniform Cartesian coordinates with $\Delta z_i$ denoting the cell width in the leakage direction, and $\Delta x_j$ denoting the frequency bin width:
\begin{equation} \label{eq:z-leakage}
  k_{z\text{-leak}}^{\delta i} = \frac{1}{3 \Delta z_i} \frac{2}{k_{i,j} \Delta z_i + k_{\delta i,j} \Delta z_{\delta i}}
\end{equation}
and
\begin{equation} \label{eq:x-leakage}
  k_{x\text{-leak}}^{\delta j} = \frac{1}{\Delta x_j} \frac{1}{k_{i,j}^{-1} \Delta x_j + k_{i,\delta j}^{-1} \Delta x_{\delta j}} \, .
\end{equation}
Here, the cell- and bin-averaged scattering coefficients are defined by $k_{i,j} \equiv (\Delta V_i \Delta x_j)^{-1}\iint_{i,j} k_x\, \text{d}V \text{d}x$. We note that the discretization is not required to follow a particular grid structure. For example, it would also be natural to have a frequency weighting kernel, such as splines for higher order convergence. In this paper we employ a tophat filter to simulate a piecewise constant frequency representation. Specifically, the Ly$\alpha$ absorption coefficient is $k_x = k_0\,H(a,x)$, where $k_0$ denotes the value at line centre and the Hjerting--Voigt function is defined in equation~(\ref{eq:H}) with a second order expansion in $a$ of \citep[see][]{Smith_2015}
\begin{equation} \label{eq:H_approx}
  H(a,x) \approx e^{-x^2} + \frac{2 a}{\sqrt{\upi}} \big(2 x\,F(x) - 1 \big) + a^2 e^{-x^2} \big(1 - 2 x^2\big) \, ,
\end{equation}
where the Dawson integral is $F(x) = \int_0^x e^{y^2-x^2} \text{d}y$. Integrating equation~(\ref{eq:H_approx}) over frequency yields
\begin{align}
  \mathcal{H}(a,x) &\equiv \int_0^x H(a,y)\,\text{d}y \notag \\
  &\approx \frac{\sqrt{\upi}}{2} \text{erf}(x) - \frac{2 a}{\sqrt{\upi}} F(x) + a^2\,x\,e^{-x^2} \, ,
\end{align}
where the error function is $\text{erf}(x) \equiv 2 \int_0^x e^{-y^2} \text{d}y / \sqrt{\upi}$. Thus, a tophat discretization has a mean absorption coefficient of
\begin{equation}
  k_{j} = k_{0} \big(\mathcal{H}(a,x_j+\Delta x_j/2) - \mathcal{H}(a,x_j-\Delta x_j/2)\big) / \Delta x_j \, ,
\end{equation}
where for clarity we have dropped the spatial index and introduced the frequency bin centre and width as $x_j$ and $\Delta x_j$, respectively.

To our best knowledge, the leakage treatment for frequency under the Fokker-Planck approximation for resonant scattering has not appeared previously in the literature. The derivation is similar to spatial leakage in 1D non-uniform Cartesian coordinates. We note that in an isothermal, uniform medium with uniform spatial and frequency meshes equation~(\ref{eq:z-leakage}) reduces to $k_{z\text{-leak}}^{\delta i} = [3 k_j (\Delta z)^2]^{-1}$. Likewise, equation~(\ref{eq:x-leakage}) reduces to $k_{x\text{-leak}}^{\delta j} = [(k_j^{-1} + k_{\delta j}^{-1})\,(\Delta x)^2]^{-1}$. Notice this is  a harmonic mean between frequency bins. Therefore, the spatial and frequency dependence may be factored out for an efficient on-the-fly implementation, with the additional advantage that the redistribution function is no longer required.

\subsection{Monte Carlo procedure}
The Ly$\alpha$ Monte Carlo procedures used to solve equation~(\ref{eq:RTE-discrete-double-diffusion}) are similar to the continuous implementation of MCRT \citep[see e.g.][]{Dijkstra_2014}. However, instead of following continuous photon trajectories, the DDMC packets are tracked by the cell index and frequency bin. When precise positions and frequencies are needed they can be drawn uniformly from the cell volume or bin interval. After a DDMC packet is initialized, the subsequent trajectory is determined by the locally dominant opacity. In other words, the leakage process with the smallest interaction distance:
\begin{equation} \label{eq:l_min_1}
  \Delta \ell = \min_{\delta i,\delta j}\left(\Delta \ell_{z\text{-leak}}^{\delta i}, \Delta \ell_{x\text{-leak}}^{\delta j}\right) \, ,
\end{equation}
is executed to transport the photon packet across the appropriate cell or frequency bin interface.
Here, the indices $\delta i$ and $\delta j$ run over all neighboring cells and frequency bins. The respective lengths are determined by drawing the effective optical depths from exponential distributions, such that
\begin{equation} \label{eq:l_min_2}
  \Delta \ell_X = -\ln \xi / k_X \qquad \text{for} \quad k_X \in \left\{k_{z\text{-leak}}^{\delta i}, k_{x\text{-leak}}^{\delta j}\right\} \, ,
\end{equation}
where $\xi$ is a random number uniformly distributed on $[0,1]$, and the label $X$ specifies the transport process. We retain physical time-dependence of the radiation field by limiting the cumulative path length of each photon packet during a timestep. Specifically, causality requires that if $\Delta \ell_X > c (t_{n+1} - t_k)$, where $t_k$ denotes the current photon time and $t_{n+1}$ the end of the timestep, then we set $t_k = t_{n+1}$ and the photon packet remains within the cell and frequency bin without executing leakage.

Photon packets are removed from the simulation when they reach the boundary of the domain; otherwise they persist across timesteps. If the number of active photons exceeds a certain value then a photon packet resampling scheme may be desirable. However, we do not implement resampling in this paper. We instead focus on the scenario in which all photons are emitted at the beginning of the simulation, or continuously as prescribed for a radiation source. For example, a time-dependent source with luminosity $L_\alpha(t)$ emits an energy of $\mathcal{E}_n = \int_n L_\alpha(t) \text{d}t$, with packet emission times uniformly distributed over the duration of the timestep, $t_k \in [t_n,t_{n+1}]$. For multiple sources, such as stellar populations or extended recombination emission, and time-variable sources, it may also be desirable to incorporate a photon packet weighting scheme to accelerate the convergence of Monte Carlo sampling.
The weights allow simulations to dynamically allocate photon packets to optimally sample the radiation field.
The total energy emitted during each timestep is distributed with weights of $w_k$ across $\Delta N_{\text{ph},n}$ photon packets, e.g. if equal weights are chosen then $w_k = \Delta N_{\text{ph},n}^{-1}$. The individual packet energies are given by $\varepsilon_k = w_k \mathcal{E}_n$, so by construction $\sum \varepsilon_k = \mathcal{E}_n$. Special emission cases include an initial flash of energy $\mathcal{E}_0$ and/or a constant luminosity source $L_\alpha$.

We employ two different Monte Carlo estimators for the spectral energy density. First, the `bin' estimate is based on counting the number of photons within each cell and bin at the end of each timestep. Second, the more accurate `path' estimate is based on the total residence time due to the continuous propagation of photon packets \citep{Smith_RHD_2017}. The spectral energy density estimators are thus given by
\begin{equation} \label{eq:E_estimator}
  E_{i,j,n} \approx \sum_\text{bins} \frac{\varepsilon_k}{\Delta V_i \Delta x_j} \approx \sum_\text{paths} \frac{\varepsilon_k}{\Delta V_i \Delta x_j} \frac{\Delta\ell_k}{c \Delta t_n} \, ,
\end{equation}
where the summations are over all photon packets within the particular cell, frequency bin, and timestep. We calculate the integrated energy density via $E_{i,n} \approx \sum_j E_{i,j,n} \Delta x_j$ for frequency and $E_{j,n} \approx \sum_i E_{i,j,n} \Delta V_i$ for space. Similarly, a Monte Carlo estimator for the acceleration of a given cell under the DDMC transport scheme is
\begin{equation} \label{eq:a_estimator}
  \bmath{a}_{i,n} \approx \sum_\text{faces} \frac{\varepsilon_k \bmath{\tau}_{\delta i,j}}{c \rho_i \Delta t_n} \, ,
\end{equation}
where the summation is over all leaked photon packets, inward and outward, across all faces of the current cell throughout the timestep. If the outward unit normal vector of the cell faces are denoted by $\bmath{n}_{\delta i}$, then the optical depth vector denotes the contribution along the leakage direction according to $\bmath{\tau}_{\delta i,j} \equiv \pm k_{i,j} \Delta \ell_{z\text{-leak}}^{\delta i} \bmath{n}_{\delta i}$, where a positive (negative) sign denotes outward (inward) leakage. As with energy density, it is also possible to retain the frequency information; however, for clarity we do not do that here.

\subsection{Dust absorption}
The absorption of Ly$\alpha$ photons by dust can have a significant impact on the internal radiation field and emergent flux. We briefly describe two methods to account for dust absorption in the DDMC scheme. In both cases the leakage coefficients are calculated using the total scattering coefficient including Ly$\alpha$ and dust, such that $k_{i,j} = k_{\alpha,i,j} + k_\text{s}$ in equations (\ref{eq:z-leakage}) and (\ref{eq:x-leakage}), where dust scattering is typically modeled assuming a constant albedo defined by $A \equiv k_\text{s}/(k_\text{s}+k_\text{a})$. The first method follows a probabilistic treatment by sampling the distance to an absorption event as $\Delta \ell_\text{a} = -\ln \xi / k_\text{a}$, similar to equation~(\ref{eq:l_min_2}). If this is smaller than the traversed leakage distance from equation~(\ref{eq:l_min_1}) then the photon packet is eliminated from the simulation after a residence time within the cell of $\Delta t_\text{a} = \Delta \ell_\text{a} / c$. The second method employs continuous absorption by adjusting the weight of photon packets to accelerate the convergence of the radiation field. In other words, if the photon packet moves with an interaction distance of $\Delta \ell$ based on equation~(\ref{eq:l_min_1}), then the traversed optical depth is $\tau_\text{a} = k_\text{a} \Delta \ell$ and the weight is reduced by $e^{-\tau_\text{a}}$. Photon packets with negligible weight can be eliminated by applying a threshold condition, e.g. $w_k > w_\text{min} \sim 10^{-10}$. We note that in this scheme the Monte Carlo estimators in equations (\ref{eq:E_estimator}) and (\ref{eq:a_estimator}) must be modified to account for the decreasing contribution along the path. Otherwise the momentum transfer would be overestimated, especially in cases where the cumulative optical depth to escape is greater than unity, i.e. $\sum \tau_\text{a} > 1$. The correction factor to the path estimators for energy density and acceleration before applying the weight reduction is $\int_0^{\tau_\text{a}} e^{-\tau'_\text{a}}\,\text{d}\tau'_\text{a}/\tau_\text{a} = (1 - e^{-\tau_\text{a}}) / \tau_\text{a}$, which is approximately $1 - \tau_\text{a}/2$ in the optically-thin limit. In this paper we only consider dust-free settings.

\subsection{Comoving frame formulation}
Bulk velocity can strongly affect Ly$\alpha$ radiative transfer as the resulting Doppler shifts can bring photons closer to or further from resonance. Ly$\alpha$ scattering is most naturally calculated in the gas comoving frame, typically in the non-relativistic limit. Therefore, we incorporate bulk velocity with a Doppler frequency correction $\Delta\nu$ when moving across cell interfaces,
\begin{equation}
  \frac{c \Delta \nu}{\nu_0} = -\bmath{n}_k \cdot \Delta \bmath{v} \, .
\end{equation}
The corresponding dimensionless frequency correction introduces a factor accounting for temperature variations across cells, $x' = (x + \bmath{n}_k \cdot \bmath{u}) \sqrt{T/T'} - \bmath{n}_k \cdot \bmath{u}'$, where the primed quantities pertain to the new cell and $\bmath{u} \equiv \bmath{v} / v_\text{th}$. For discrete spatial transport the net Doppler shift only depends on the flux directionality function $W(\mu)$, averaged over outward-facing directions. Thus, the DDMC change in velocity is
\begin{equation} \label{eq:velocity-shift}
  \frac{c \Delta \nu}{\nu_0} = -f_\mu \Delta v_{\delta i} \qquad \text{where} \quad f_\mu \equiv \frac{\int_0^1 \mu W(\mu)\,\text{d}\mu}{\int_0^1 W(\mu)\,\text{d}\mu} \, .
\end{equation}
Here, $\Delta v_{\delta i}$ denotes the magnitude of the velocity difference in the direction normal to the $\delta i$ neighboring cell interface. The coefficient $f_\mu$ quantifies the assumed anisotropy at the interface. For reference, in Doppler units equation~(\ref{eq:velocity-shift}) becomes $x' = x \sqrt{T/T'} + f_\mu (\Delta u_{\delta i} \sqrt{T/T'} - \Delta u_{\delta i}')$. In this paper we employ the approximation of isotropic escape for pure DDMC transport such that $W(\mu) \propto 1$ and $f_\mu = 1/2$, which should be considered as a conservative lower limit. We note that if $W(\mu) \propto \mu$ then $f_\mu = 2/3$, and if $W(\mu) \propto \mu^2$ then $f_\mu = 3/4$. Therefore, if focusing does occur for photons emerging from planar boundaries into lower density gas then we still expect $f_\mu \lesssim 0.7$. This averaging scheme can also independently incorporate discrete frequency bins. We simply apply the linear correction to the bin edges and draw a new random frequency uniformly from within the Doppler-shifted range. We then reassign the DDMC packet to the corresponding discrete frequency bin.

\section{Test problems}
\label{sec:tests}
To validate the accuracy of our resonant DDMC scheme we implement a suite of test problems. Where possible we derive analytic solutions for comparison to the numerical calculations. The setup for grey diffusion in Section~\ref{sec:grey-diffusion} validates the spatial DDMC scheme. Then in Section~\ref{sec:static-medium-resonant-scattering} we investigate the behavior of Ly$\alpha$ scattering in spherical and plane-parallel geometries. We present new approximate steady-state solutions for an infinite, static, homogeneous medium. We also compare the evolution of a Ly$\alpha$ radiation pulse with a pulse evolving under grey diffusion.

\begin{figure}
  \centering
  \includegraphics[width=\columnwidth]{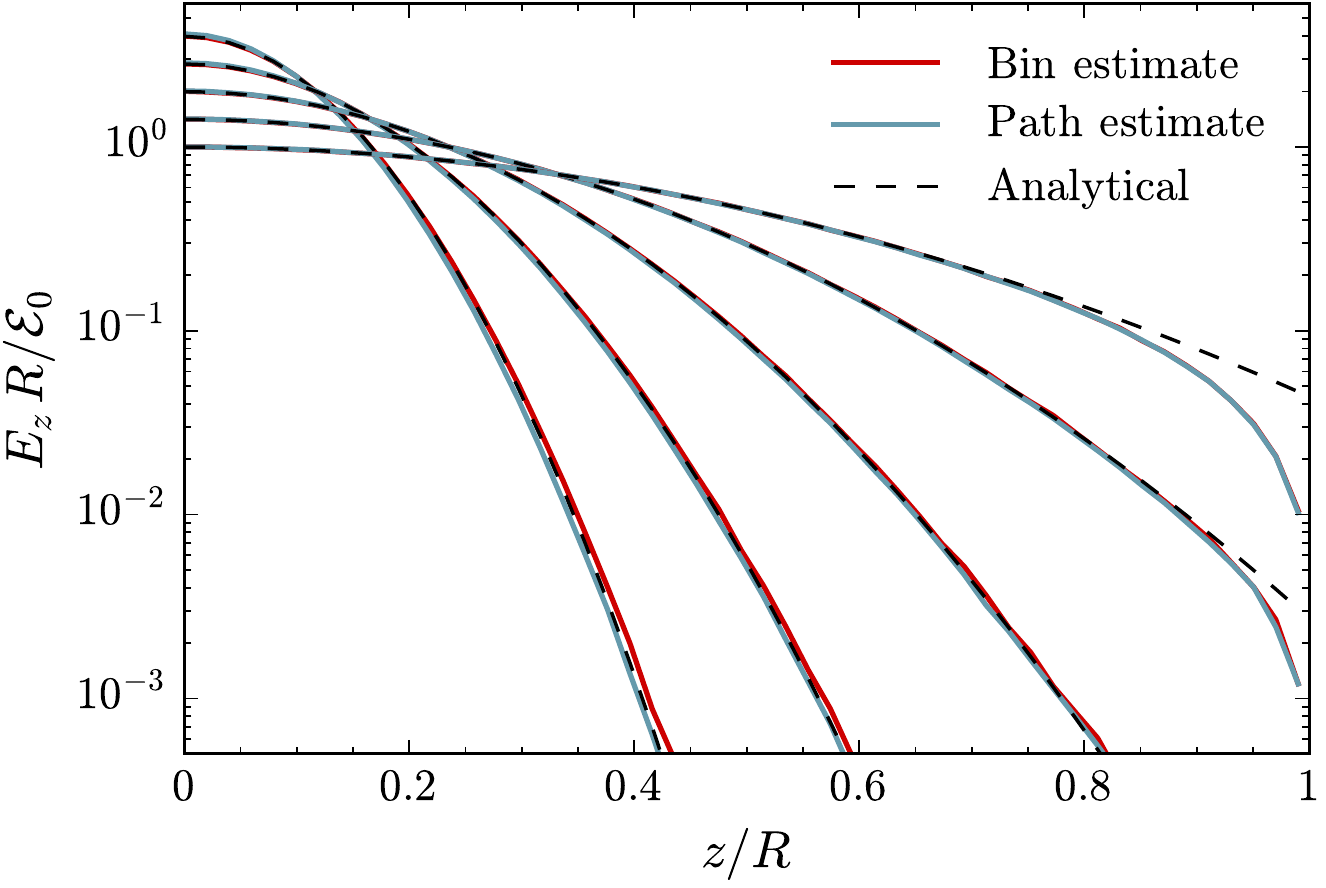}
  \caption{Radiation energy density $E_z$ for grey diffusion in a uniform medium over several doubling times, $t = \{1, 2, 4, 8, 16\} \times 10^{-2}\,t_\text{diff}$, where the diffusion time is $t_\text{diff} = \frac{1}{2} \tau t_\text{light} = k R^2 / 2 c$. For convenience the axes have been rescaled into dimensionless units. Both photon `bin' counting and cumulative `path' length Monte Carlo estimators are shown as described by equation~(\ref{eq:E_estimator}). Within the domain, both provide excellent agreement with the analytical solution of equation~(\ref{eq:RTE-grey-solution}).}
  \label{fig:grey}
\end{figure}

\subsection{Grey diffusion}
\label{sec:grey-diffusion}
We consider the spatial transport of DDMC particles in a 1D infinite domain. We restrict to the case of a constant scattering coefficient, which is equivalent to a random walk in space with a uniform mean free path of $\lambda_\text{mfp} = k^{-1}$. Under these conditions the time-dependent diffusion equation is
\begin{equation} \label{eq:RTE-grey}
  \frac{1}{c} \frac{\partial E_z}{\partial t} = \frac{1}{k} \frac{\partial^2 E_z}{\partial z^2} + \frac{\mathcal{E}_0}{c} \delta(z) \delta(t) \, ,
\end{equation}
where the notation for the source function fully specifies the initial value problem, i.e. $E_z|_{t=0^+} = \mathcal{E}_0 \delta(z) / c$. For an arbitrary reference length scale $R$ the light crossing time is $t_\text{light} = R/c$ and the diffusion time is $t_\text{diff} = \frac{1}{2} \tau t_\text{light} = k R^2 / 2 c$. Rescaling equation~(\ref{eq:RTE-grey}) into dimensionless units according to $z = \tilde{z}\,R$, $E_z = \tilde{E}_{\tilde{z}}\,\mathcal{E}_0 / R$, and $t = \tilde{t}\,k\,R^2 / 2 c$ yields
\begin{equation} \label{eq:RTE-grey-tilde}
  \frac{\partial \tilde{E}_{\tilde{z}}}{\partial \tilde{t}} = \frac{1}{2} \frac{\partial^2 \tilde{E}_{\tilde{z}}}{\partial \tilde{z}^2} + \delta(\tilde{z}) \delta(\tilde{t}) \, ,
\end{equation}
which has the solution:
\begin{equation} \label{eq:RTE-grey-solution}
  \tilde{E}_{\tilde{z}} = \frac{e^{-\tilde{z}^2 / 2 \tilde{t}}}{\sqrt{2 \upi \tilde{t}}} \, ,
\end{equation}
Our spatial transport test consists of a 1D grid with DDMC activated in all of the $101$ equally-spaced cells. We initialize $10^7$ DDMC particles at $t = 0$ and employ constant time steps of $\Delta t = 10^{-3} t_\text{diff}$. Figure~\ref{fig:grey} shows the radiation energy density profile over several doubling times, specifically $t = \{1, 2, 4, 8, 16\} \times 10^{-2}\,t_\text{diff}$. The simulation provides excellent agreement with the exact analytical solution of equation~(\ref{eq:RTE-grey-solution}). However, we note that the simulation domain is finite with $z \in [-R,R]$ so the two solutions diverge as photons escape when they reach boundary.

\begin{figure}
  \centering
  \includegraphics[width=\columnwidth]{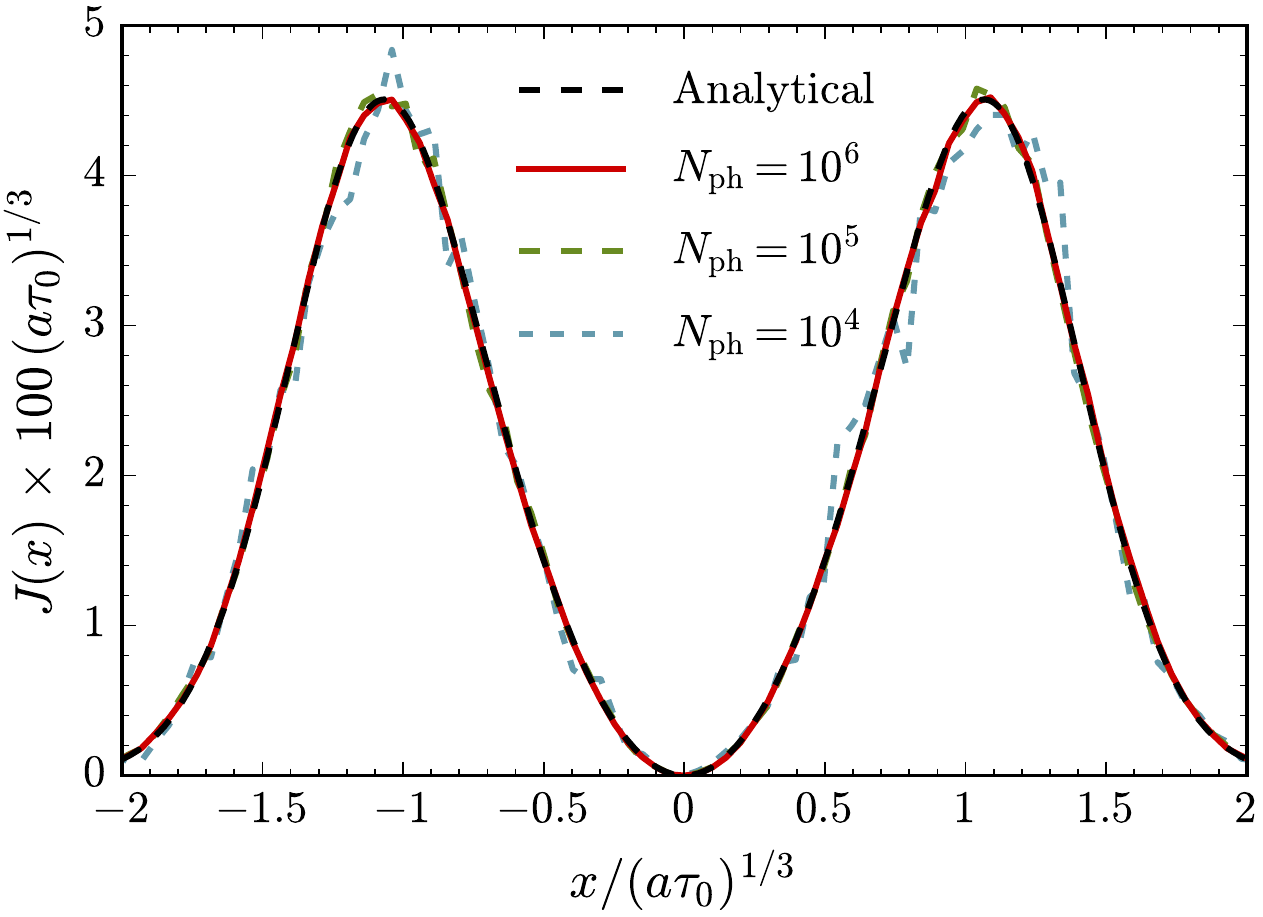}
  \caption{Angular-averaged flux $J(x)$ for a homogeneous, static slab as a function of frequency. The axes have been rescaled by factors of $(a \tau_0)^{1/3}$ to remove the dependence on temperature and optical depth from the analytic solution. The simulations were run with $a \tau_0 = 10^9$ and $T = 10$\,K. With a uniform resolution of $\Delta z \approx 0.01\,R$ and $\Delta x \approx 0.05\,(a \tau_0)^{1/3}$, the Monte Carlo noise is still apparent when the number of photon packets is $N_\text{ph} \approx 10^4$. Convergence is obtained by increasing $N_\text{ph}$, yielding nearly exact agreement to the analytic solution with $N_\text{ph} \approx 10^6$.}
  \label{fig:slab}
\end{figure}

\subsection{Static medium with resonant scattering}
\label{sec:static-medium-resonant-scattering}
We now explore the time-dependent solution for the equivalent 3D line transfer problem. The evolution is more complicated due to the diffusion in both space and frequency. The general equation assuming a static, homogeneous medium with a frequency-dependent scattering coefficient $k(x)$ is
\begin{equation} \label{eq:RTE-Lya}
  \frac{1}{c} \frac{\partial E}{\partial t} = \frac{\nabla^2 E}{3 k(x)} + \frac{\partial}{\partial x} \left( \frac{k(x)}{2} \frac{\partial E}{\partial x} \right) + \frac{\mathcal{E}_0}{c} \delta(\bmath{r}) \delta(x) \delta(t) \, ,
\end{equation}
where for notational compactness we have dropped the subscripts on $E$, and the source term specifies the initial value problem, i.e. $E_{\bmath{r},x}|_{t=0^+} = \mathcal{E}_0 \delta(\bmath{r}) \delta(x) / c$. Rescaling equation~(\ref{eq:RTE-Lya}) into dimensionless units with an arbitrary reference length scale $R$ is accomplished by $\bmath{r} = \tilde{\bmath{r}}\,R$, $E = \tilde{E}\,\mathcal{E}_0 / R^3$, $\text{d}x = \sqrt{3/2}\,k(x) R\,\text{d}\tilde{x}$, and $t = \tilde{t}\,3 R / c$, yields
\begin{equation} \label{eq:RTE-Lya-tilde}
  \tau({\tilde{x}}) \frac{\partial \tilde{E}}{\partial \tilde{t}} = \tilde{\nabla}^2 \tilde{E} + \frac{\partial^2 \tilde{E}}{\partial \tilde{x}^2} + \sqrt{\frac{2}{3}}\,\delta(\tilde{\bmath{r}}) \delta(\tilde{x}) \delta(\tilde{t}) \, ,
\end{equation}
where the optical depth is defined as $\tau(\tilde{x}) = k(\tilde{x}) R$.

\begin{figure}
  \centering
  \includegraphics[width=\columnwidth]{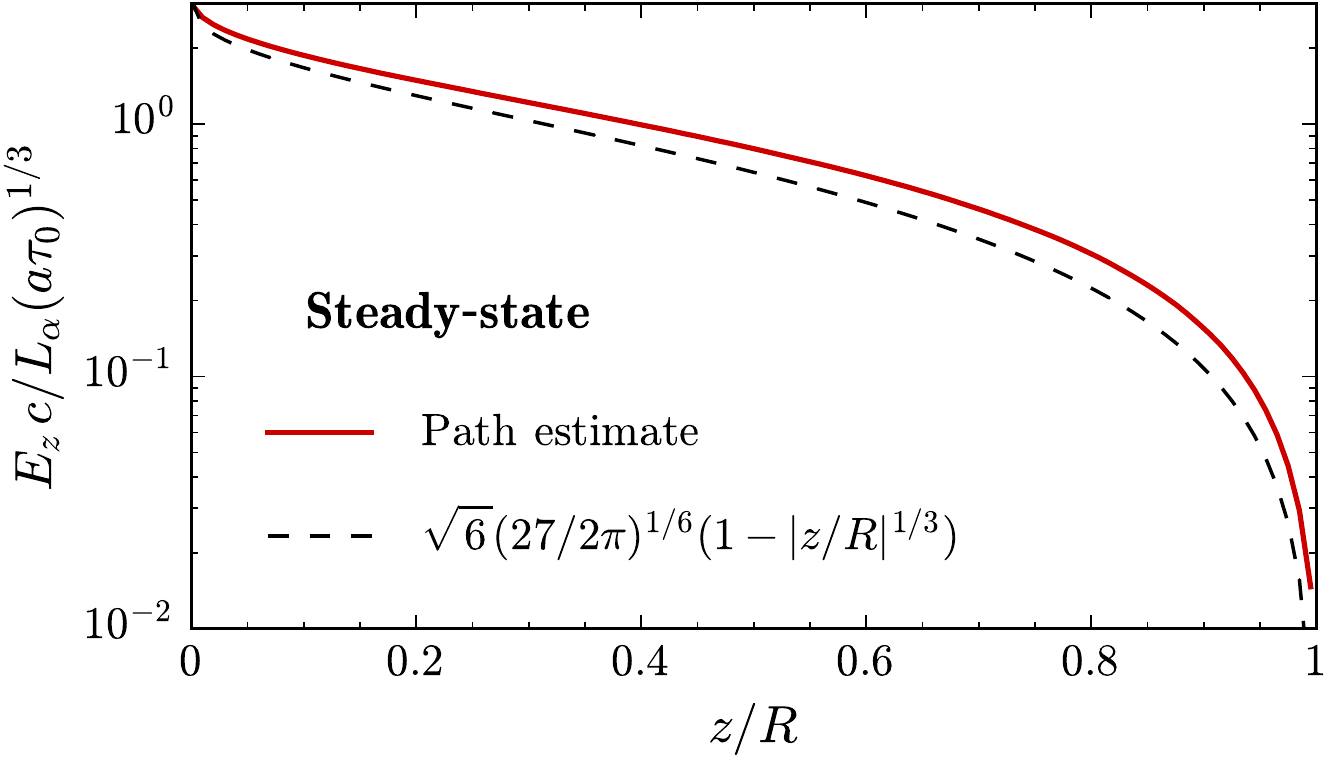}
  \caption{Radiation energy density $E_z$ for the steady-state case of resonant scattering in a static uniform medium. For convenience the axes have been rescaled into dimensionless units. The cumulative path length Monte Carlo estimator is shown along with the approximate analytical solution of equation~(\ref{eq:steady-state-solution-full-slab-z}), which is within the expected agreement (see text).}
  \label{fig:steady-state-slab-z}
\end{figure}

\subsubsection{Steady-state}
We first consider a steady-state point source by eliminating the time derivative in equation~(\ref{eq:RTE-Lya-tilde}) and by replacing the source term with $\mathcal{E}_0 \delta(t) \rightarrow L_\alpha$ and the dimensionless transformation with $E = \tilde{E}\,L_\alpha / c R^2$. The spatial and frequency terms may be combined to take advantage of 4D spherical symmetry by defining $\xi \equiv \sqrt{\tilde{r}^2 + \tilde{x}^2}$, where $\tilde{r} = \| \tilde{\bmath{r}} \|$, further simplifying equation~(\ref{eq:RTE-Lya-tilde}) into
\begin{equation} \label{eq:RTE-Lya-tilde-steady-state}
  \frac{\partial}{\partial \xi} \left( \xi^3 \frac{\partial \tilde{E}}{\partial \xi} \right) = -\frac{\sqrt{6}}{2 \upi^2} \delta(\xi) \, .
\end{equation}
The source term has a factor of $\sqrt{6}$ due to the coefficients of $1/3$ and $1/2$ in equation~(\ref{eq:RTE-Lya}), and $2 \upi^2$ in the denominator is the surface area of the 4D unit sphere. Equation~(\ref{eq:RTE-Lya-tilde-steady-state}) has solution
\begin{equation} \label{eq:steady-state-solution}
  \tilde{E}(\xi) = \frac{\sqrt{6}}{4 \upi^2} \xi^{-2} \, .
\end{equation}
which may also be derived by relating the symmetric 4D Fourier transform to the equivalent 1D Hankel transform. The final form after substituting back the original spatial and frequency variables involves an integral over the frequency-dependent optical depth. The integral can be evaluated analytically for wing photons, i.e. $|x| \gtrsim x_\text{cw}$, using equation~(\ref{eq:H})
\begin{equation} \label{eq:x-tilde}
  \tilde{x} \sqrt{\frac{3}{2}} = \int \tau_x^{-1} \text{d}x \approx \int \frac{\sqrt{\upi} x^2}{a \tau_0} \text{d}x = \frac{\sqrt{\upi} x^3}{3 a \tau_0} \, .
\end{equation}
Therefore, after converting to physical units we have:
\begin{equation} \label{eq:steady-state-solution-full}
  E(r,x) \approx \frac{\sqrt{6}}{4 \upi^2}\frac{L_\alpha}{c R^2} \left[ \left(\frac{r}{R}\right)^2 + \left( \sqrt{\frac{2 \upi}{27}} \frac{x^3}{a \tau_0}\right)^2 \right]^{-1} \, .
\end{equation}
At the characteristic radius $R$ the profile of equation~(\ref{eq:steady-state-solution-full}) has a mean of zero, absolute mean of $\langle |x| \rangle_R = (a \tau_0 / \sqrt{2 \upi})^{1/3} \approx 0.7362\,(a \tau_0)^{1/3}$, and standard deviation of $(\langle x^2 \rangle - \langle x \rangle^2)^{1/2}_R = (3 \sqrt{3} a \tau_0 / 4 \sqrt{\upi})^{1/3} \approx 0.9016\,(a \tau_0)^{1/3}$. An approximate radial energy density for wing photons may be obtained by integrating equation~(\ref{eq:steady-state-solution-full}) over frequency:
\begin{equation} \label{eq:steady-state-solution-full-r}
  E(r) \approx \frac{L_\alpha}{2 \upi c} \left( \frac{2 a \tau_0}{\sqrt{\upi} r^5 R} \right)^{1/3} \, .
\end{equation}
Likewise, the radially-integrated spectral energy density is
\begin{equation} \label{eq:steady-state-solution-full-x}
  E(x) \approx \frac{9 L_\alpha}{8 \upi c R} \frac{a \tau_0}{\sqrt{\upi} |x|^3} \, .
\end{equation}
From these simple expressions we can derive insight about Ly$\alpha$ trapping near point sources, when $r \lesssim R$. For example, the trapping time as a function of radius is roughly
\begin{align} \label{eq:t_trap}
  t_\text{trap} &\approx 4 \upi \int_0^R \frac{E(r)}{L_\alpha} r^2 \text{d}r \notag \\
  &\approx \frac{3}{2} \left( \frac{2 a \tau_0}{\sqrt{\upi}} \right)^{1/3} t_\text{light} \notag \\
  &\approx 21.94 \; t_\text{light} \, \left( \frac{N_\text{\HI}}{10^{20}\,\text{cm}^{-2}} \right)^{1/3} \left( \frac{T}{10^4\,\text{K}} \right)^{-1/3} \notag \\
  &\approx 12.14 \; t_\text{light} \, \left( \frac{\tau_0}{10^6} \right)^{1/3} \left( \frac{T}{10^4\,\text{K}} \right)^{-1/6} \, ,
\end{align}
in excellent agreement with previous estimates and empirical calculations \citep{Adams_1975,Smith_RHD_2017}.

\begin{figure}
  \centering
  \includegraphics[width=\columnwidth]{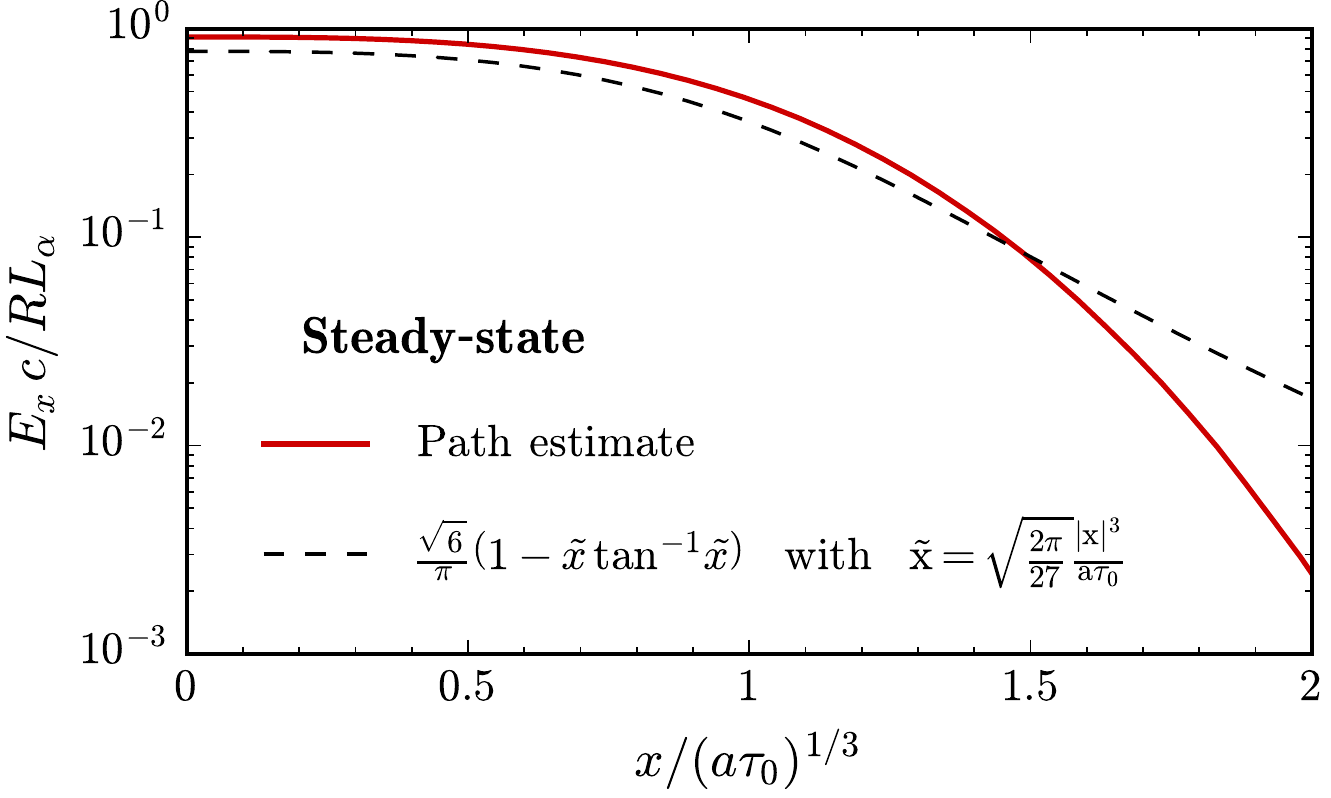}
  \caption{Spatially integrated spectral energy density $E_x$ for the same simulation as Fig.~\ref{fig:steady-state-slab-z}. The approximate analytical solution of equation~(\ref{eq:steady-state-solution-full-slab-x}) is within the expected agreement (see text).}
  \label{fig:steady-state-slab-x}
\end{figure}

However, if the optically-thick slab has finite extent then the angular averaged intensity $J_x$ at the surface is described by the following solution derived by \citet{Harrington_1973} and \citet{Neufeld_1990} \citep[see also][]{Dijkstra_2006}:
\begin{equation} \label{eq:Neufeld}
  J(x, a \tau_0) = \frac{1}{4\sqrt{6 \upi}} \frac{x^2}{a \tau_0} \text{sech} \left( \sqrt{\frac{\upi^3}{54}} \frac{x^3}{a \tau_0} \right) \, ,
\end{equation}
which has been normalized to $1/4 \upi$, reflecting an integration over solid angle. Equation~(\ref{eq:Neufeld}) is a symmetric, double-peaked profile with peaks located at $x_\text{p} = \pm 1.0664 \, (a \tau_0)^{1/3}$, which is derived by setting $\partial J / \partial x = 0$ to obtain the transcendental relation $\bar{x} \tanh \bar{x} = 2 / 3$ with $\bar{x} = \sqrt{\upi^3/54} \, x^3 / a \tau_0$. Therefore, the peak heights correspond to $10^2 \, J(\tau_0, a, x_\text{p}) = 4.5074 \, (a \tau_0)^{-1/3}$. Figure~\ref{fig:slab} illustrates the emergent Ly$\alpha$ line flux after scaling out the dependence of $(a \tau_0)^{1/3}$ from equation~(\ref{eq:Neufeld}). The simulation was run with $a \tau_0 = 10^9$ and $T = 10$\,K, with $10^6$ photon packets at a uniform resolution of $\Delta z \approx 0.01\,R$ and $\Delta x \approx 0.05\,(a \tau_0)^{1/3}$. The DDMC simulation provides excellent agreement with the analytical steady state solution in the high-opacity limit.

\begin{figure}
  \centering
  \includegraphics[width=\columnwidth]{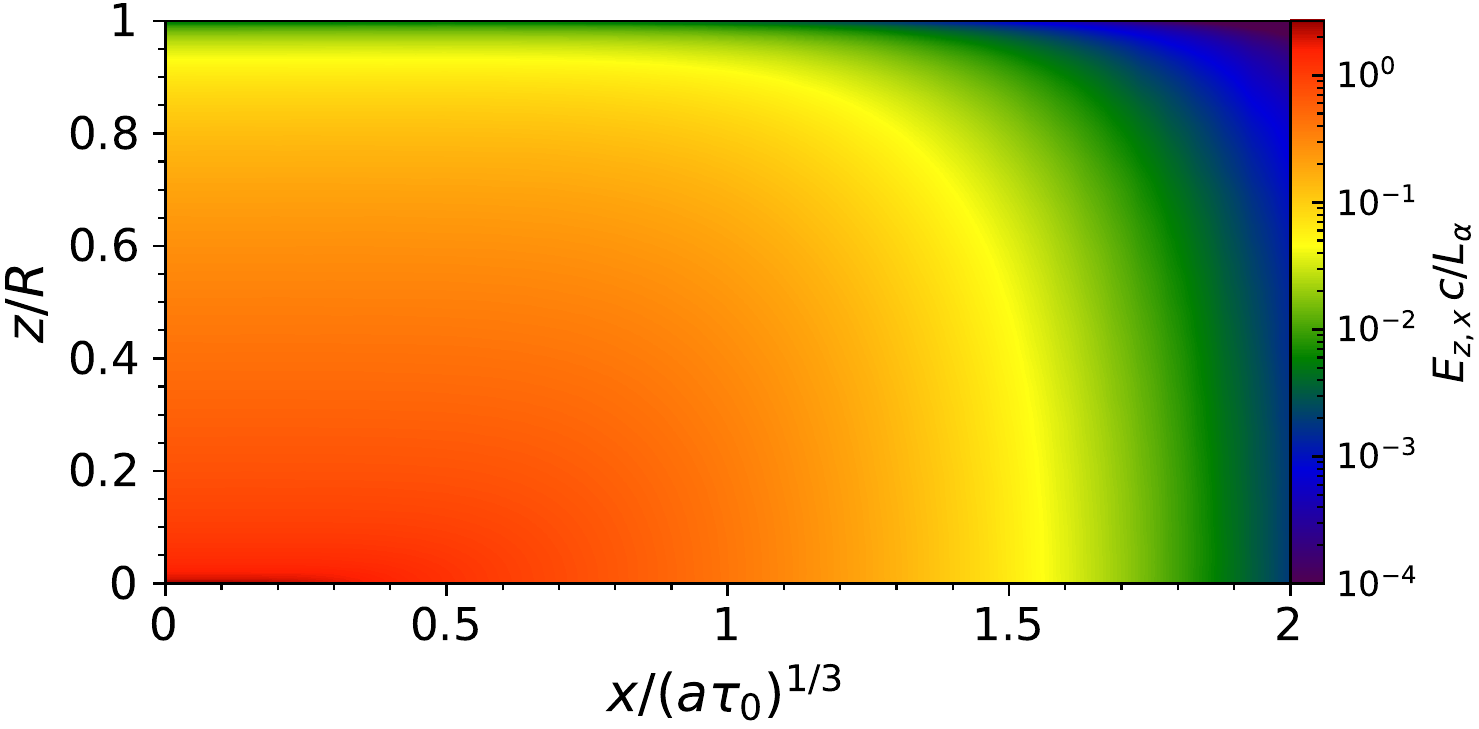}
  \caption{Spectral energy density corresponding to Figs.~\ref{fig:steady-state-slab-z} and \ref{fig:steady-state-slab-x}. The double-peaked profile is only apparent in the escaping line flux as in Fig.~\ref{fig:slab}. Otherwise, the steady-state source retains photons near line-centre as approximately described by equation~(\ref{eq:steady-state-solution-full-slab}).}
  \label{fig:steady-state-slab-both}
\end{figure}

We now compare the steady-state DDMC calculation with the analytical solution. The previous derivation must be slightly modified to account for the slab geometry. The final expression for the spectral energy density as a function of space and frequency, analogous to equation~(\ref{eq:steady-state-solution-full}), is
\begin{equation} \label{eq:steady-state-solution-full-slab}
  E(z,x) \approx -\frac{\sqrt{6}}{4 \upi}\frac{L_\alpha}{c} \log\left[ \left(\frac{z}{R}\right)^2 + \left( \sqrt{\frac{2 \upi}{27}} \frac{x^3}{a \tau_0}\right)^2 \right] \, .
\end{equation}
Unfortunately, in one dimension further integration over the infinite space and frequency domains does not converge, unless we restrict the range. For presentational simplicity we limit the respective ranges to $|z| < R$ and $|x| < \frac{\upi}{3} (\sqrt{27/2 \upi} a \tau_0)^{1/3}$. The frequency integration yields a hypergeometric function, however we obtain a simple expression from the leading order Taylor expansion. This process leads to an approximate slab version of equation~(\ref{eq:steady-state-solution-full-r}) of
\begin{equation} \label{eq:steady-state-solution-full-slab-z}
  E(z) \approx \frac{\sqrt{6} L_\alpha}{c} \left( \sqrt{\frac{27}{2 \upi}} a \tau_0 \right)^{1/3} \left( 1 - \left| \frac{z}{R} \right|^{1/3} \right) \, .
\end{equation}
The corresponding spatial integration leads to a leading order approximate slab version of equation~(\ref{eq:steady-state-solution-full-x}) of
\begin{equation} \label{eq:steady-state-solution-full-slab-x}
  E(x) \approx \frac{\sqrt{6} R L_\alpha}{\upi c} \left[ 1 - \sqrt{\frac{2 \upi}{27}} \frac{|x|^3}{a \tau_0} \arctan\left(\sqrt{\frac{2 \upi}{27}} \frac{|x|^3}{a \tau_0}\right) \right] \, .
\end{equation}
We note that in this approximation the line profile is broader than in the radial case, e.g. at the characteristic radius $R$ the profile of equation~(\ref{eq:steady-state-solution-full-slab}) has a mean of zero and standard deviation of $(\langle x^2 \rangle - \langle x \rangle^2)^{1/2}_R \approx 1.161\,(a \tau_0)^{1/3}$.

These simplified expressions allow a direct comparison with the numerical solutions obtained with the DDMC method. Figures~\ref{fig:steady-state-slab-z} and \ref{fig:steady-state-slab-x} show the energy density $E_z$ and the spatially integrated spectral energy density $E_x$ for the same steady-state simulation as Fig.~\ref{fig:slab}. The full spatial and frequency dependence from the simulation is illustrated in Fig.~\ref{fig:steady-state-slab-both}. For the steady-state case, there is no double-peaked frequency profile interior to the slab, even though this is the distinguishing characteristic of the escaping line flux as shown in Fig.~\ref{fig:slab}. The simulation and the approximate analytical solutions are within the expected agreement considering the underlying assumptions. For example, in the finite domain we expect edge effects near the slab boundary as well as for wing photons that have previously escaped. In the 1D case the analytic expressions are based on a restricted integration range that cannot disentangle correlation between position and frequency, which may be important for a finite slab. The apparent discrepancy may also be due to the assumed line opacity, which is the full Voigt profile in the simulation instead of the wing approximation made in the analytic expressions. Finally, the pure DDMC implementation may slightly overestimate the path length for free-streaming photons. In any case, the simplicity of the analytical solutions and the rough agreement with simulations demonstrate the utility of this kind of analysis for Ly$\alpha$ transfer calculations.

\begin{figure}
  \centering
  \includegraphics[width=\columnwidth]{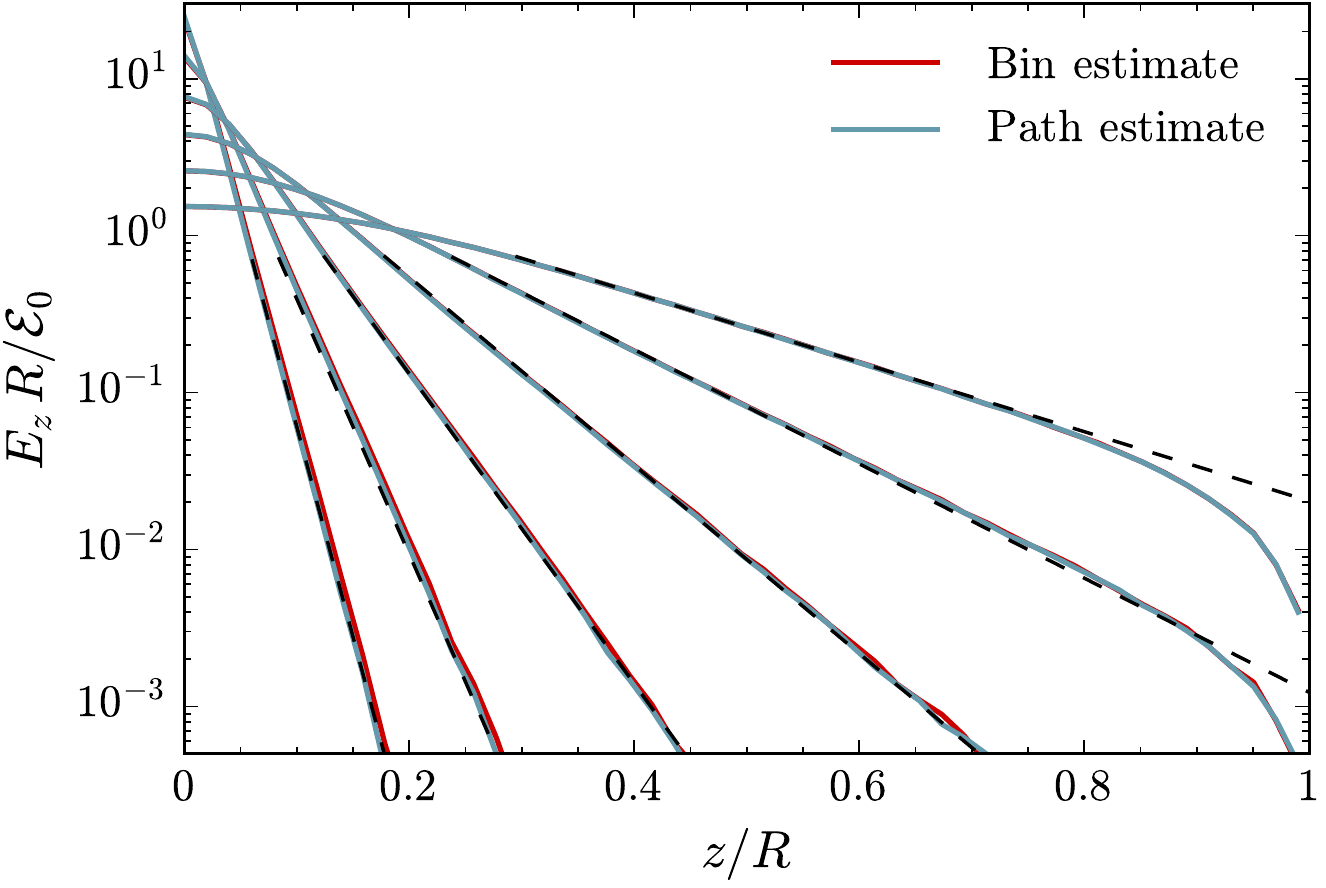}
  \caption{Radiation energy density $E_z$ for resonant scattering in a static uniform medium over several doubling times, $t = \{1, 2, 4, 8, 16, 32\} \times 10^{-2}\,(a \tau_0)^{1/3} R / c$. For convenience the axes have been rescaled into dimensionless units. Both photon bin counting and cumulative path length Monte Carlo estimators are shown. The main difference from grey diffusion is the shape of the spatial evolution (see Fig.~\ref{fig:grey}).}
  \label{fig:time-slab-z}
\end{figure}

\begin{figure}
  \centering
  \includegraphics[width=\columnwidth]{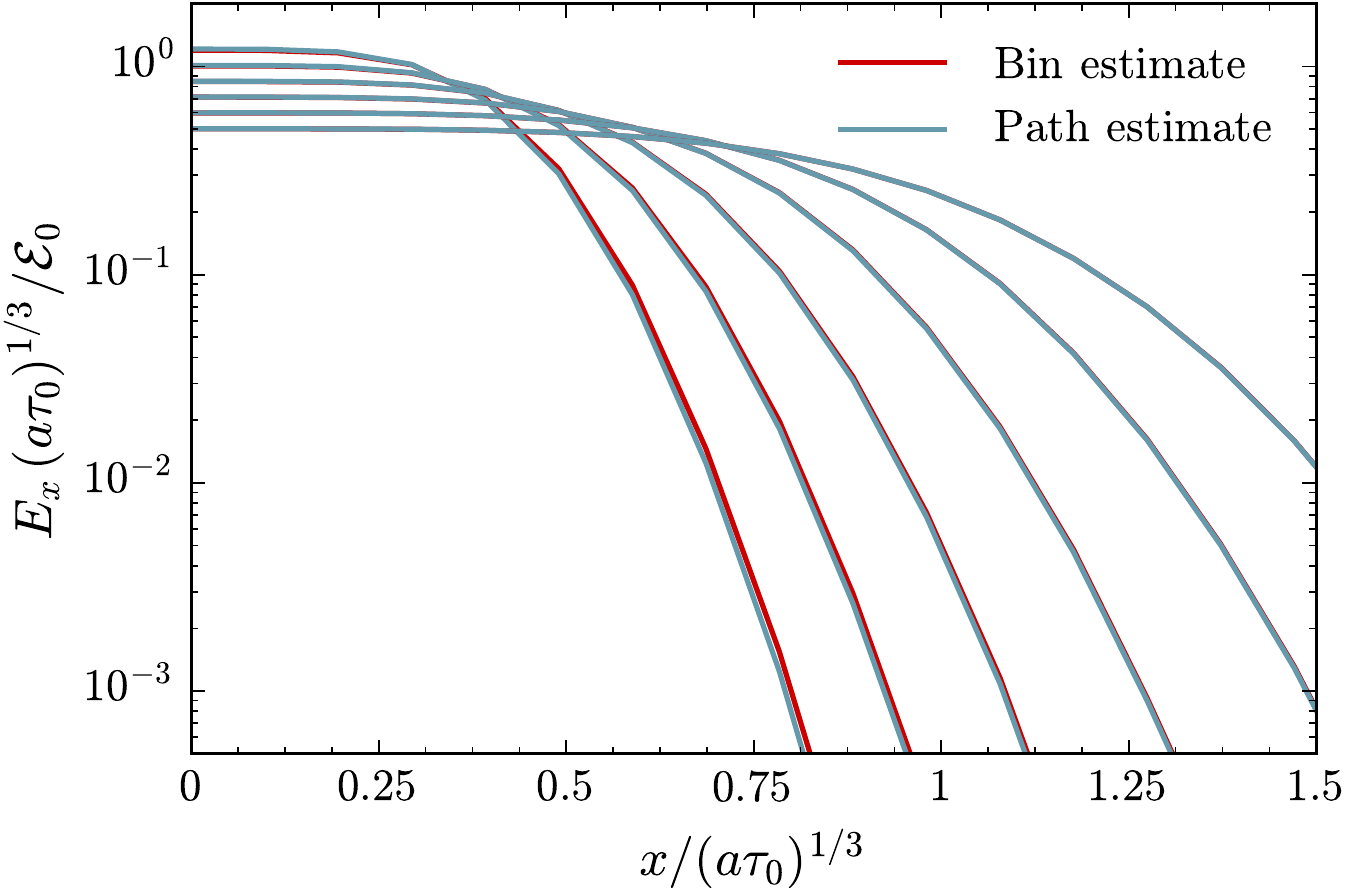}
  \caption{Spatially integrated spectral energy density $E_x$ for the same simulation as illustrated in Fig.~\ref{fig:time-slab-z}. The distribution gradually spreads out and flattens with time.}
  \label{fig:time-slab-x}
\end{figure}

\subsubsection{Time-dependence}
\begin{figure}
  \centering
  \includegraphics[width=\columnwidth]{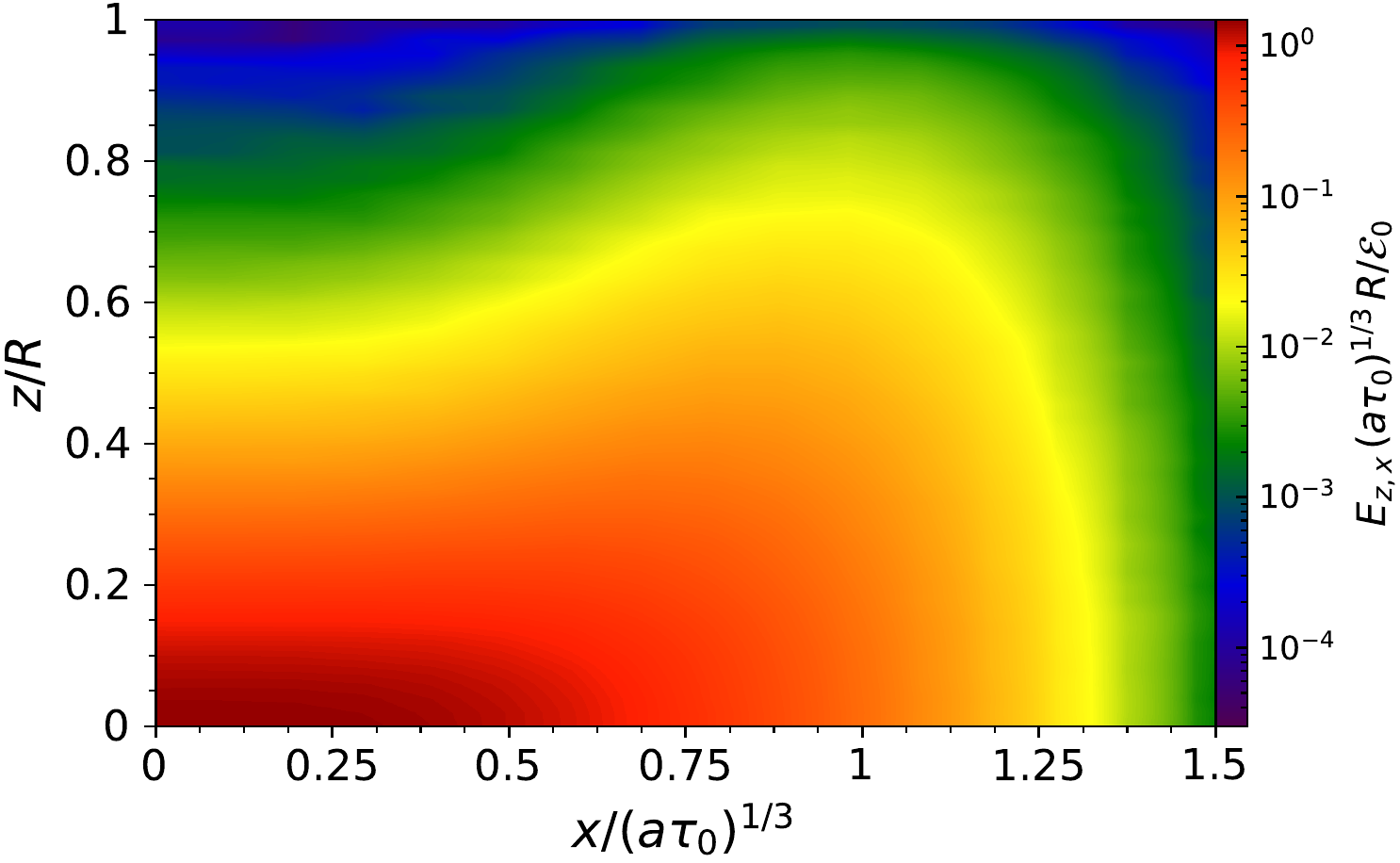}
  \caption{Spectral energy density corresponding to Figs.~\ref{fig:time-slab-z} and \ref{fig:time-slab-x} at the time $t = 0.2\,(a \tau_0)^{1/3} R / c$. The double-peaked profile is already apparent near the boundary.}
  \label{fig:time-slab-both}
\end{figure}

We also explore the time-dependent solution of equation~(\ref{eq:RTE-Lya-tilde}), describing the evolution of a pulse of radiation undergoing diffusion in space and frequency. For testing purposes the DDMC scheme is activated in all cells and frequency bins. We initialize the packets at $t = 0$ and employ a constant time step of $\Delta t = 10^{-3} (a \tau_0)^{1/3} R / c$, where we choose $a \tau_0 = 10^9$ with $T = 10$\,K. For reasons discussed in Section~\ref{sec:Lya-performance} this particular test has uniform cells of width $\Delta z \approx 0.02\,R$ and uniform bins of width $\Delta x \approx 0.1\,(a \tau_0)^{1/3}$. The simulation was run with $10^7$ photon packets. Figures~\ref{fig:time-slab-z} and \ref{fig:time-slab-x} show the radiation energy density over several doubling times, specifically $t = \{1, 2, 4, 8, 16, 32\} \times 10^{-2}\,(a \tau_0)^{1/3} R / c$. For clarity, the profiles have been integrated over frequency in Fig.~\ref{fig:time-slab-z} and space in Fig.~\ref{fig:time-slab-x}. The evolution differs from that of grey diffusion. The spatial profile is approximately an exponential function rather than the Gaussian in grey diffusion (compare with Fig.~\ref{fig:grey}). Specifically, we find that the solution can be approximated by the fitting function $\text{d}\log E_z / \text{d}|z/R| \approx -2.24\,(t/t_\text{diff})^{-0.72}$, where the approximate diffusion time is $t_\text{diff} = t_\text{light} (a \tau_0)^{1/3}$. The spatially-averaged line profile becomes increasingly flat with time. However, the emergent profile does develop the expected double peak structure. This is apparent when considering the energy density near the boundary. Fig.~\ref{fig:time-slab-both} illustrates the nontrivial dependence on position and frequency at the time $t = 0.2\,(a \tau_0)^{1/3} R / c$.

\section{Performance Characteristics}
\label{sec:performance}
We now analyze the numerical performance of the DDMC method.  We characterize its scaling and compare its runtimes to those of counterpart non-diffusive MCRT simulations. As before we start with grey diffusion before we turn to resonant scattering.

\subsection{Grey diffusion}
In grey diffusion the discrete grid spacing acts as an effective mean free path for photon propagation, such that $\lambda_\text{mfp,eff} = \max(\Delta z, k^{-1})$. If the computational bottleneck is physical scattering, then the number of scattering events after one diffusion time is $N_\text{scat}^\text{MC} \approx c k t_\text{diff} = \tau^2 / 2$ and $N_\text{scat}^\text{DDMC} \approx c t_\text{diff} / \Delta z = \tau R / 2 \Delta z$ for the MC and DDMC schemes, respectively. Therefore, the estimated speedup by employing the DDMC method is approximately $N_\text{scat}^\text{MC} / N_\text{scat}^\text{DDMC} \approx \tau \Delta z / R$. The actual speedups may vary depending on details about the code implementation and hardware architecture. We have checked that this relation roughly holds for pure DDMC scattering across reasonable parameter ranges in $\tau$ and $\Delta z / R$. However, to illustrate the $\tau$ and $\Delta z$ performance characteristics we run a suite of simulations with varying optical depth and grid resolution. Figure~\ref{fig:grey-light} shows the relative computational time to run simulations with $\tau \in [1, 10^6]$ for one light crossing time $t_\text{light}$, while varying the number of grid cells such that $n_z = 2 R / \Delta z \in 2^{\{-\infty, 3, 6, 9\}} + 1 = \{1, 9, 65, 513\}$. With a high enough opacity, the time to diffuse out of one cell becomes longer than the overall light crossing time, so the runtime flattens off as additional speedup is no longer possible. This begins to occur when $t_\text{diff,cell} \gtrsim t_\text{light}$, a condition that can be reformulated in terms of the overall optical depth and number of cells as $\tau \gtrsim n_z^2 / 2$.

\subsection{Resonant scattering}
\label{sec:Lya-performance}
We now perform a suite of simulations to assess how the resolution affects the performance of the pure resonant DDMC scheme. Although the following is an oversimplification, the local diffusion into neighboring cells limits the frequency-dependent effective mean free path $\lambda_\text{mfp,eff} = \max(\Delta z, k_x^{-1})$. The typical redistribution in the wings generates an average drift back toward the core of $\langle \delta x \rangle \sim -1 / x$ and a RMS displacement of $\sqrt{\langle\delta x^2\rangle} \sim 1$ \citep{Osterbrock_1962}. Therefore, a wing photon tends to return to the core after $N_\text{scat,wing} \sim x^2$ scatterings. In the optically-thick regime, escape occurs after an excursion in the wing to a characteristic escape frequency determined by setting the RMS displacement to the slab size, i.e. $N_\text{scat,wing}^{1/2} \lambda_\text{mfp} \approx R$. Therefore, the escape frequency is approximately $x_\text{esc} \approx (a \tau_0 / \sqrt{\upi})^{1/3}$, the optical depth at this frequency is $\tau_\text{esc} \approx x_\text{esc}$, the mean free path is $\lambda_\text{mfp} \approx R / x_\text{esc}$, and the trapping (or diffusion) time is $t_\text{trap} \approx x_\text{esc} t_\text{light}$ \citep{Adams_1975}. However, the total number of scatterings including core photons with short mean free paths is higher and may be estimated from the cumulative escape probability via $N_\text{scat} \sim P_\text{esc}^{-1} \sim [2 \int_{x_\text{esc}}^\infty \text{d}x \phi(x) / x^2]^{-1} \sim \tau_0$ \citep{Adams_1972}. Physical scattering dominates the computational time for Ly$\alpha$ MCRT codes, so we expect the runtime to scale as $t_\text{MC} \propto \tau_0$. With a core-skipping acceleration scheme this is improved to $t_\text{MC,cs} \propto x_\text{esc}^2 \propto (a \tau_0)^{2/3}$. We have verified that this is approximately true in our MCRT runs.

\begin{figure}
  \centering
  \includegraphics[width=\columnwidth]{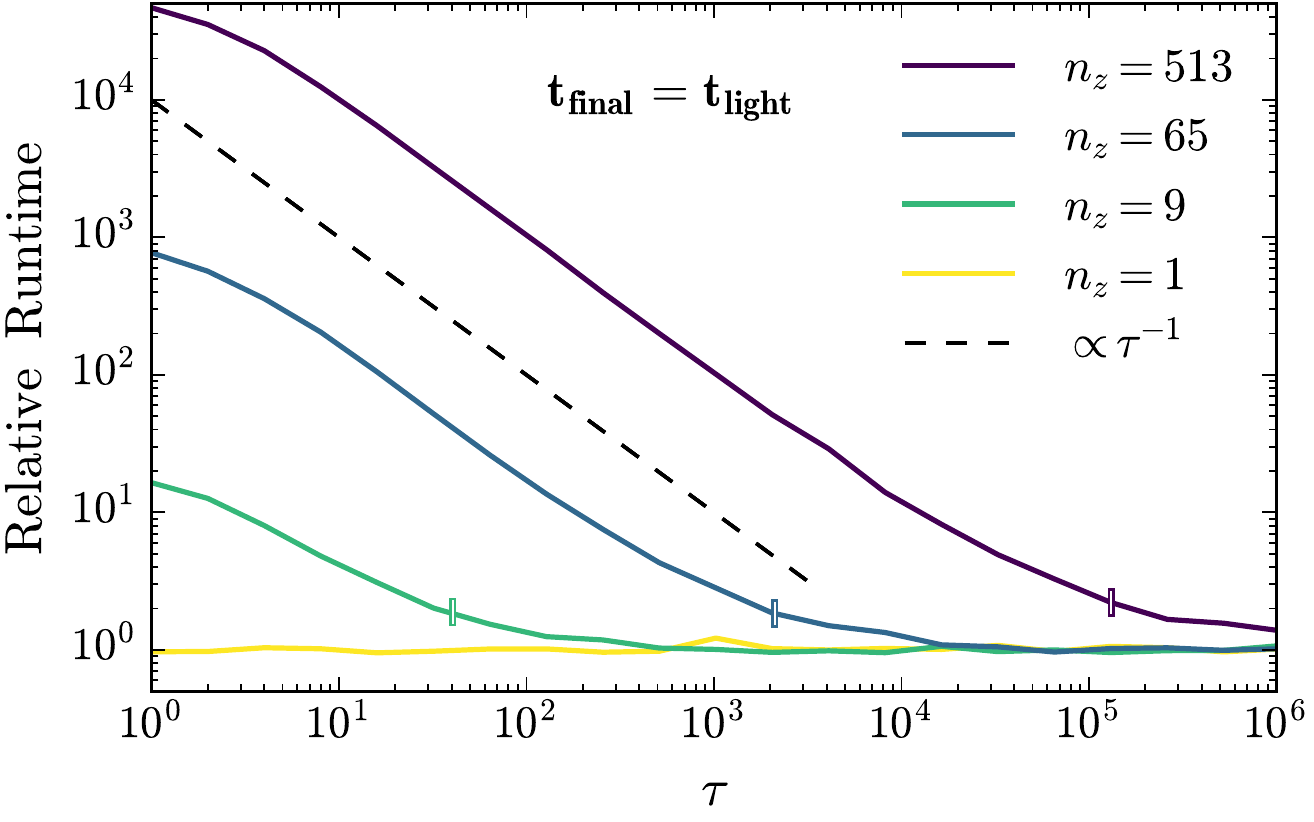}
  \caption{Relative runtime for DDMC scattering simulations carried out for one light crossing time as a function of optical depth $\tau$. The number of grid cells $n_z = 2 L / \Delta z \in 2^{\{-\infty, 3, 6, 9\}} + 1 = \{1, 9, 65, 513\}$. The flattening occurs when the diffusion time across one cell becomes longer than the light crossing time of the box, i.e. $\tau \gtrsim n_z^2 / 2$, marked by vertical ticks for each relevant curve. At lower optical depths the relative runtime is inversely proportional to opacity.}
  \label{fig:grey-light}
\end{figure}

\begin{figure}
  \centering
  \includegraphics[width=\columnwidth]{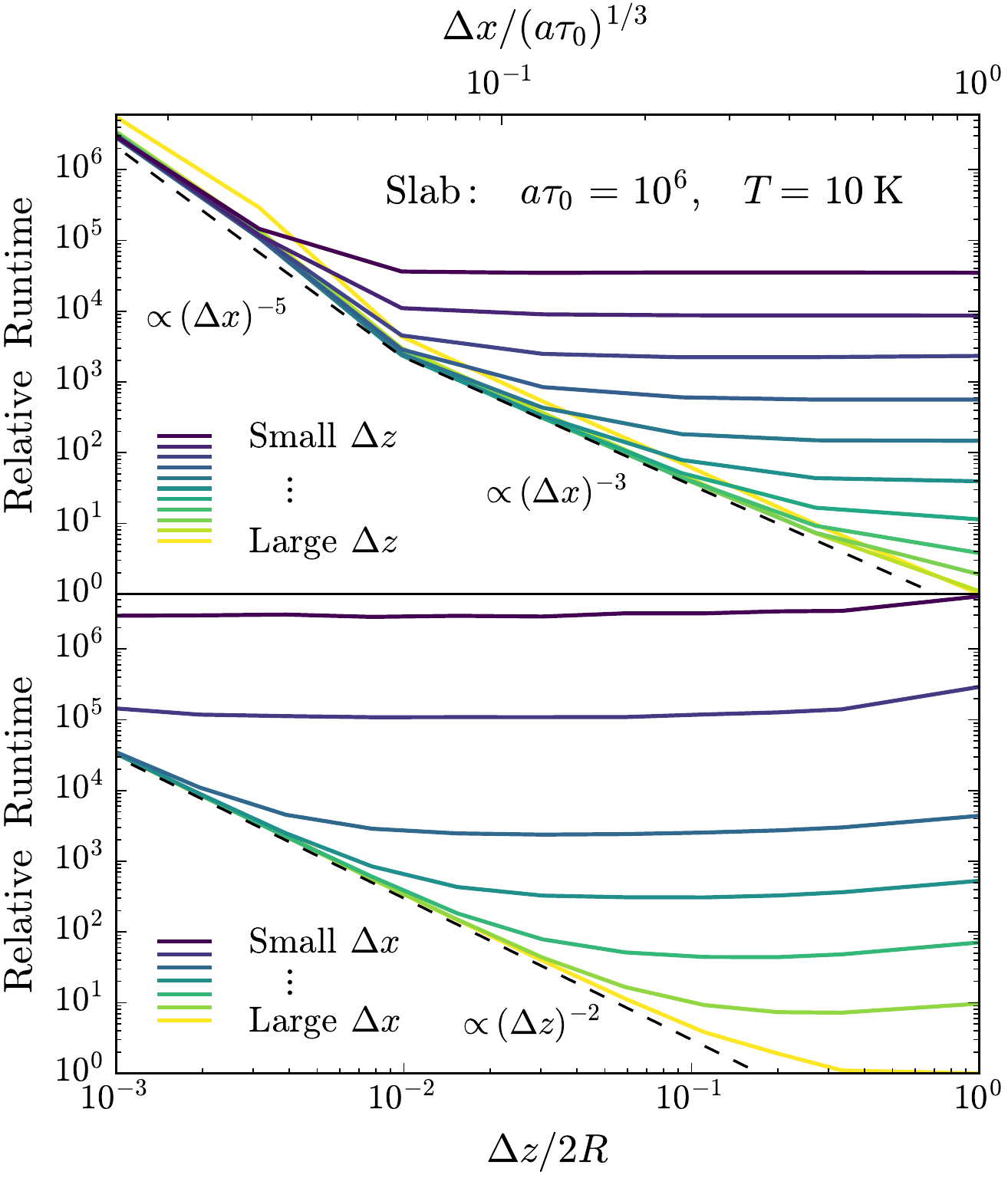}
  \caption{Relative runtimes for a suite of simulations with varying discretization for the spatial cells and frequency bins. The runtimes are normalized to the most efficient DDMC simulation. The simulations finish when all photon packets escape a slab with $a \tau_0 = 10^6$ and $T = 10$\,K. The upper (lower) panel illustrates the parameter space as a function of bin (cell) size. The runtime scales as $t_\text{DDMC} \propto (\Delta x)^{-3}$ when frequency diffusion dominates and $t_\text{DDMC} \propto (\Delta z)^{-2}$ when spatial diffusion dominates. Otherwise, the timing flattens out as the other discretization resolution (cell or bin) acts as the computational bottleneck. The apparent break in the power law in the upper panel corresponds to a bin size of $\Delta x \sim 2 x_\text{cw} \approx 5$, which implies less efficient diffusive redistribution for core photons.}
  \label{fig:slab-timing-1d}
\end{figure}

On the other hand, the runtime for escape under a pure DDMC implementation is independent of the optical depth but is determined by the spatial and frequency discretization. However if the comparison is across a physical timescale such as the light crossing time then the relative DDMC runtime may decrease with increasing opacity, similar to the trend demonstrated in Fig.~\ref{fig:grey-light}. For concreteness we test the pure DDMC scheme with different cell and bin sizes. The simulations finish when all photon packets have escaped the slab with $a \tau_0 = 10^6$ at a temperature of $T = 10$\,K. In this case, the wing dominated transport calculations scale as $t_\text{DDMC} \propto (\Delta z)^{-2}$ and $t_\text{DDMC} \propto (\Delta x)^{-3}$, as demonstrated in Fig.~\ref{fig:slab-timing-1d}. This is due to physical scatterings being replaced by leakage events. The spatial term is the usual diffusion process with a scattering reduction factor of $\sim (\Delta z / \lambda_\text{mfp})^2 \sim \tau_{x,\text{cell}}^2$.

The frequency redistribution also induces a drift term, which makes diffusion from the core to the escape frequency more difficult. This may be seen by expanding the frequency derivative in equation~(\ref{eq:RTE-continuous-double-diffusion}) as $k_x \partial^2 J/ \partial x^2 - k_x' \partial J/ \partial x$. We argue that the speedup includes a diffusive factor of $\sim \Delta x^2 /  \langle\delta x^2 \rangle$ and an additional factor $\propto \Delta x$ from the drift toward the core. Therefore, the computational efficiency may be limited by the cell size or bin width if either is too small. In this case, the two are well matched when the following condition is approximately satisfied:
\begin{equation} \label{eq:Delta_xz}
  \frac{\Delta x^3}{a \tau_0} \approx 2.5 \frac{\Delta z^2}{R^2} \, ,
\end{equation}
which explains the flattening of the timings in Fig.~\ref{fig:slab-timing-1d}.

\begin{figure}
  \centering
  \includegraphics[width=\columnwidth]{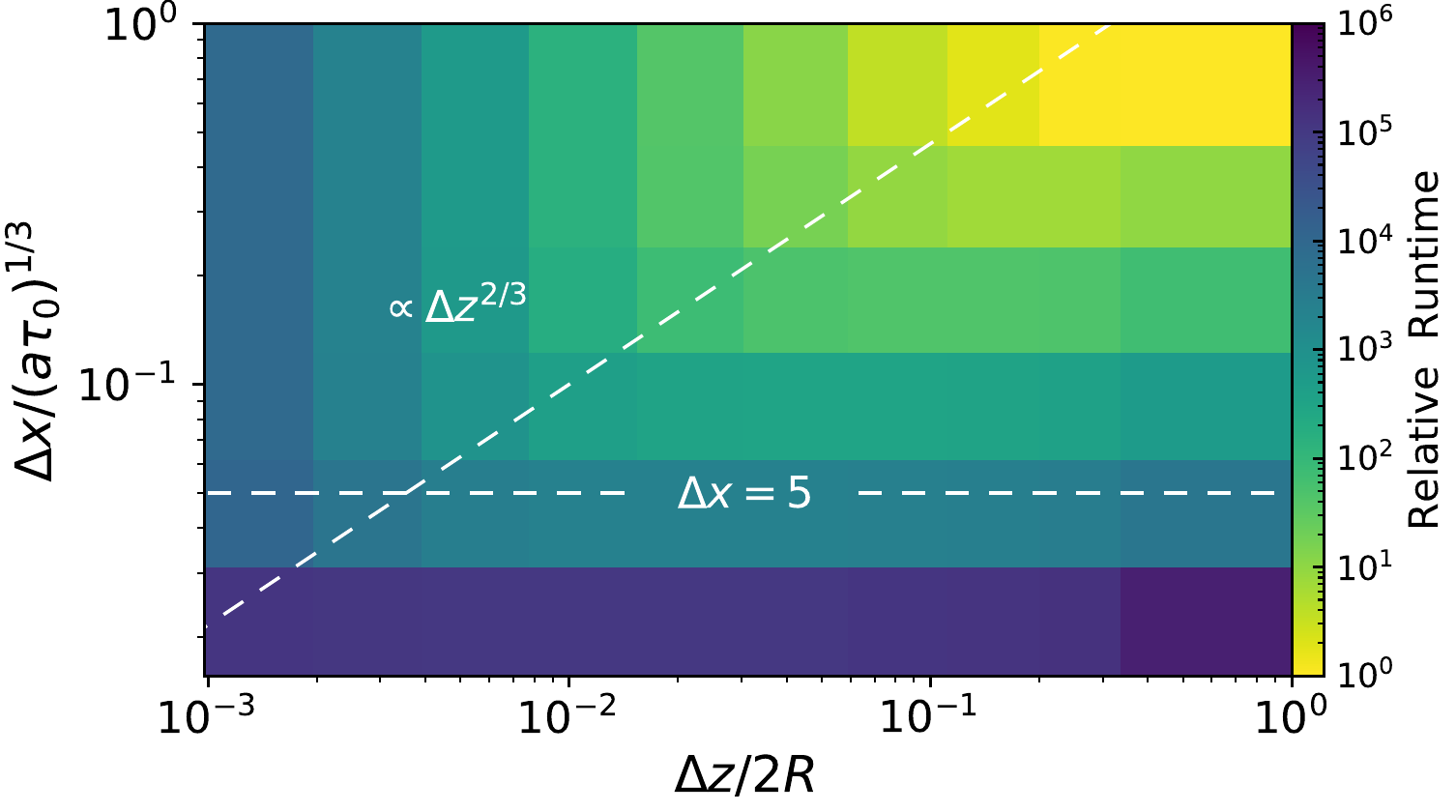}
  \caption{Relative runtime for the same suite of simulations shown in Fig.~\ref{fig:slab-timing-1d}. The main features are illustrated with the dashed lines: (\emph{i.}) an efficiency barrier at $\Delta x \lesssim 2 x_\text{cw} \approx 5$, and (\emph{ii.}) a matching effect where the cell and bin sizes yield comparable computational costs given by equation~(\ref{eq:Delta_xz}), such that $\Delta x \propto (\Delta z)^{2/3}$.}
  \label{fig:slab-timing-2d}
\end{figure}

For additional clarity, Fig.~\ref{fig:slab-timing-2d} illustrates this matching effect for the same suite of pure DDMC simulations of varying cell and bin sizes. If the computational domain covers the range $|z| \leq R$ and $|x| \leq 2.5\,(a \tau_0)^{1/3}$, then the approximate relation given by equation~(\ref{eq:Delta_xz}) corresponds to a matched number of cells and bins when $n_x^3 \approx 12.5\,n_z^2$. Therefore, a reasonable choice to obtain satisfactory resolution in test cases is $n_z = 101$ and $n_x = 51$. Finally, we note that the DDMC method is most efficient when $\Delta x$ is larger than the width of the core $\sim 2 x_\text{cw}$. Such frequency binning provides an effective core-skipping scheme because frequency redistribution within the same bin is skipped.

\subsection{Comparison with MCRT}
\label{sec:MCRT-comparison}
\begin{figure}
  \centering
  \includegraphics[width=\columnwidth]{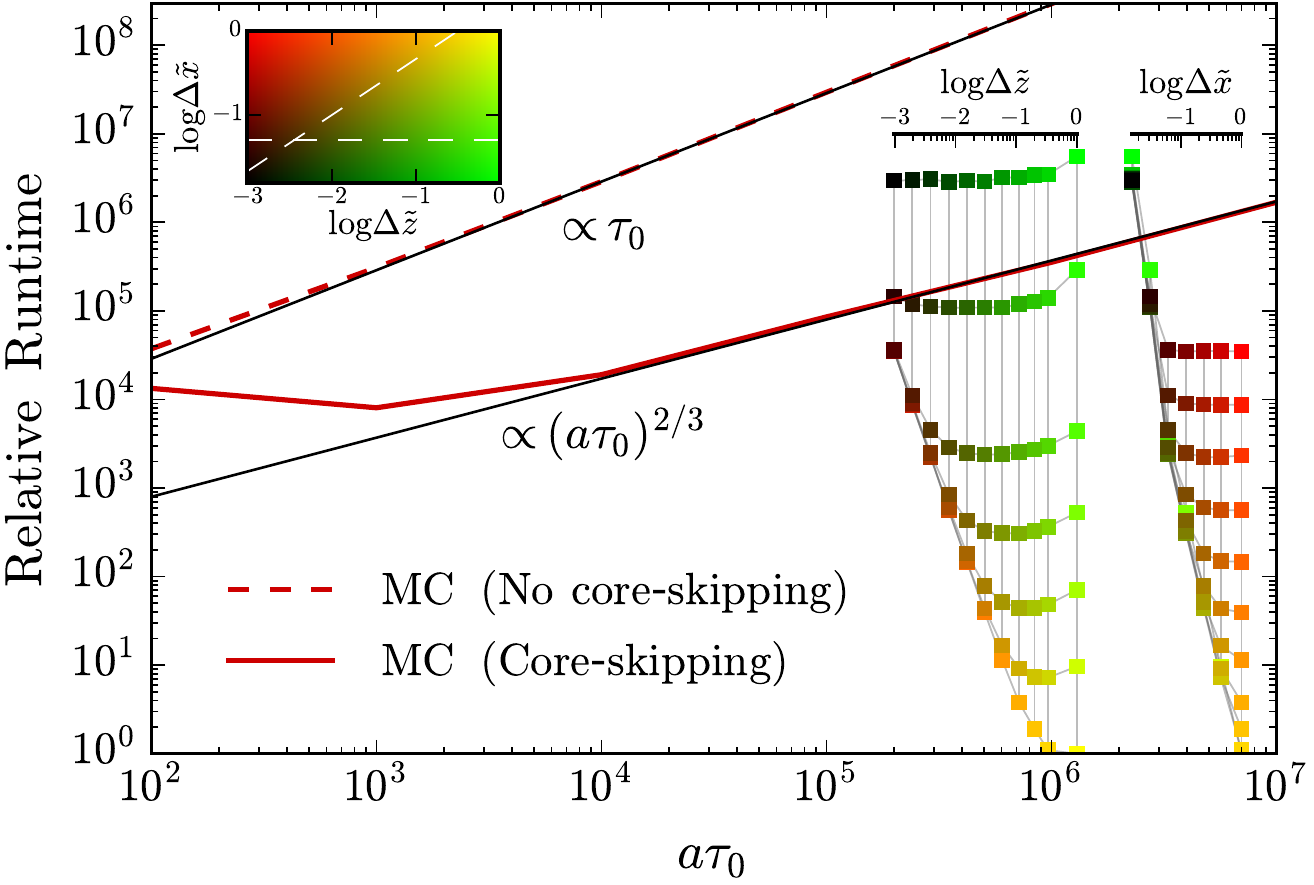}
  \caption{Relative runtimes for the DDMC and continuous MC methods with the same normalization as Figs.~\ref{fig:slab-timing-1d} and \ref{fig:slab-timing-2d}. The setup allows for direct comparison as the MC timings are from different simulations that finish when all photon packets escape a slab of $a \tau_0$ at $T = 10$\,K. The MC runtime scales with the average number of scatterings, which is $\propto \tau_0$ without the core-skipping acceleration scheme and $\propto (a \tau_0)^{2/3}$ otherwise. The runtime for a pure DDMC implementation is independent of opacity. Therefore, we show the same DDMC data from the suite of simulations at $a \tau_0 = 10^6$. For presentation purposes we insert 1D axes illustrating the cell (bin) size dependence at constant bin (cell) widths, labeled by $\tilde{z} \equiv z / 2 R$ and $\tilde{x} \equiv x / (a \tau_0)^{1/3}$, respectively. The upward rungs of the left (right) lattice show the effect of halving the bin (cell) width. For clarity, we also include a 2D color diagram, with red (green) indicating that the runtime is dominated by spatial (frequency) diffusion. The $\propto (\Delta z)^{-2}$ and $\propto (\Delta x)^{-3}$ scalings are evident in these respective regimes. The dashed lines in the color insert highlight the efficiency barrier when $\Delta x \lesssim 2 x_\text{cw}$ and the matching effect is illustrated in Fig.~\ref{fig:slab-timing-2d}.}
  \label{fig:MCRT-timings}
\end{figure}

The strength of the DDMC method is the relative computational speedup compared to continuous MC in the high opacity limit. This is because the MC runtime scales with the typical number of scatterings, which is $\propto \tau_0$ without a core-skipping acceleration scheme and $\propto (a \tau_0)^{2/3}$ with core-skipping. However, the runtime in a pure DDMC implementation depends only on the spatial cell and frequency bin discretization. For concreteness, Fig.~\ref{fig:MCRT-timings} compares DDMC and MC performance under a similar setup as in Figs.~\ref{fig:slab-timing-1d} and \ref{fig:slab-timing-2d}. Specifically, the MC timings are from different simulations that finish when all photon packets escape. The setups are characterized by the parameter $a \tau_0 = 2.7728\,(T/10\,\text{K})^{-1}\,(N_\text{H}/10^{14}\,\text{cm}^{-2})$, which for consistency is evaluated at a temperature of $T = 10$\,K. The DDMC simulation suite has $a \tau_0 = 10^6$, but with varying cell and bin sizes. The relative runtime is the physical runtime scaled by the total number of photon packets, such that the shortest time is unity. An explanation of the presentation details is given in the figure caption. The main trends have already been discussed. Still, Fig.~\ref{fig:MCRT-timings} demonstrates the potential speedup obtained by implementing rDDMC. We also provide a 2D color map with the red (green) indicating where the runtime is dominated by spatial (frequency) diffusion. In summary, our DDMC method can be several orders of magnitude more efficient than standard MCRT even without overly sacrificing on the spatial and spectral resolution of the simulation.

\section{Future directions}
\label{sec:developments}
We now discuss some future considerations to be addressed toward applying our DDMC method in general 3D Ly$\alpha$ transport codes.

\subsection{Hybrid scheme}
\label{sec:hybrid}
The DDMC method is accurate as long as the diffusion approximation holds. The approximation may be violated for wing photons even in optically-thick environments. This limitation can be resolved with a hybrid transport scheme in which DDMC packets can convert to spatially-continuous MC packets and vice versa, depending on whether the frequency-dependent optical depth of the cell is sufficiently large, e.g. how $\tau_{i,j} \equiv k_{i,j} \Delta z_i$ compares to $\tau_\text{DDMC}$. The case of spatial leakage into lower opacity cells is described by \citet{Densmore_2007}, \citet{Abdikamalov_2012}, and \citet{Tsang_2017}, who adopt the `asymptotic diffusion limit' as the interfacing boundary condition.

For completeness, we sketch out the main ideas of the hybrid scheme for \emph{spatial} transport.
When calculating the leakage coefficients we first determine whether the neighboring cells are optically thick. However, the $k_{\delta i,j}$ in equation~(\ref{eq:z-leakage}) denotes the comoving opacity of the neighbor so we must account for velocity induced Doppler shifting. If $\tau_{\delta i,j} < \tau_\text{DDMC}$ then the leakage coefficient is redefined to be
\begin{equation}
  k_{\text{DDMC}\rightarrow\text{MC}}^{\delta i} = \frac{1}{3 \Delta z_i} \frac{2}{k_{i,j} \Delta z_i + 2 \lambda} \, ,
\end{equation}
where $\lambda \approx 0.7104$ is the constant extrapolation distance \citep{Habetler_1975}. If this corresponds to the minimum distance according to equations~(\ref{eq:l_min_1})~and~(\ref{eq:l_min_2}), then the DDMC packet becomes an MC packet with a random position on the cell interface and an isotropic outward direction. On the other hand, if an MC packet moves into a neighboring cell that is optically thick then it is converted into a DDMC packet in that cell with probability
\begin{equation} \label{eq:P_MC_to_DDMC}
  P_{\text{MC}\rightarrow\text{DDMC}}^{\delta i} = \frac{2}{k_{\delta i,j} \Delta z_{\delta i} + 2 \lambda} \left( \frac{2}{3} + \mu \right) \, ,
\end{equation}
where $\mu$ is the directional cosine for the MC packet with respect to the cell interface. Otherwise, the packet scatters back into the original cell with a random position on the cell interface and an isotropic inward direction. For equation~(\ref{eq:P_MC_to_DDMC}) to have a valid probabilistic interpretation, we require $P_{\text{MC}\rightarrow\text{DDMC}}^{\delta i} \in [0,1]$ for $\mu \in (0,1]$, which imposes a condition that $\tau_\text{DDMC} \gtrsim 2$, although we suggest a more conservative choice of $\tau_\text{DDMC} = 5$. The conversion between DDMC and MC spatial schemes may also occur when a packet shifts in frequency. Therefore, after a scattering event, $\tau_{i,j}$ is recomputed and conversion is considered and executed if appropriate.

The implementation of a hybrid \emph{frequency} scheme is more subtle. Specifically, the Fokker-Planck approximation is most accurate for wing photons where the frequency redistribution is local, i.e. for $x_j \gtrsim 5$ the RMS displacement is $ \sqrt{\langle\delta x^2 \rangle} \sim 1$. On the other hand, the core photons, which undergo complete redistribution, present a challenge, but depending on the application it may not be necessary to accurately model core photons. Therefore, we suggest minimum resolution for the central frequency bin of $\Delta x_0 \gtrsim 2 x_\text{cw}$ as an effective core-skipping scheme, and $\Delta x_j \gtrsim 1$ in the wings. Interfacing may be achieved by defining an analogous frequency flux of $G_x \equiv \frac{1}{2}k_x\,\partial J_x/\partial x$ and applying appropriate boundary conditions. There may also be other ways to address the limitations of the Fokker-Planck approximation. We defer that analysis to future work.

\subsection{Continuous MCRT}
After converting a DDMC packet into a continuous MC packet, the packet $k$ is assigned additional a continuous position $\bmath{z}_k$, propagation direction $\bmath{n}_k$, and/or frequency $x_k$. In the case of continuous spatial transport, the leakage coefficient is replaced with the minimum interface-crossing distance, e.g. $\Delta \ell_k \leq \Delta z_k^{\delta i} / \mu_k^{\delta i}$, where $z_k^{\delta i}$ is the orthogonal distance from the MC packet to the interface with cell $\delta i$ and $\mu_k^{\delta i}$ is the directional cosine defined by $\bmath{n}_k$ and the interface normal vector. Whether a scattering event or interface crossing occurs, the packet position changes by
\begin{equation}
  \Delta \bmath{z}_k = \Delta \ell_k \bmath{n}_k \, .
\end{equation}
In the case of continuous frequency transport, whenever a scattering event takes place the scattering coefficient is replaced with the normal value of $k_x$. The corresponding redistribution process is exactly the same as in pure resonant MCRT. Specifically, the scattering is coherent in the rest frame of the atom, with the parallel velocity component affected by the presence of the resonance line. Thus, the change in frequency is
\begin{align}
  \Delta x_k &= \Delta \bmath{n}_k \cdot \bmath{u}_\text{atom} + g\,(\mu - 1) \notag \\
           &= (u_\parallel - g)\,(\mu - 1) + u_\perp \sqrt{1-\mu^2} \, ,
\end{align}
where $\Delta \bmath{n}_k$ denotes the vector difference between outgoing and ingoing propagation directions, $\mu$ the directional cosine between the two directions, $\bmath{u}_\text{atom}$ the atom's velocity in Doppler units, and $g \equiv h \Delta \nu_\text{D} / 2 k_\text{B} T$ the recoil parameter \citep[see e.g.][]{Adams_1971}. The second equality utilizes a special reference frame aligned with the motion of the scattering atom, such that the probability distribution functions for the parallel and perpendicular components are $P(u_\parallel) \propto \exp(u_\parallel^2) / [a^2 + (x - u_\parallel)^2]$ and $P(u_\perp) \propto \exp(u_\perp^2)$, respectively. A core-skipping acceleration scheme for frequency diffusion may also be used as in resonant MCRT \citep[e.g.][]{Smith_2015}.

\subsection{Discretized frequency redistribution}
In principle, MCRT can incorporate either continuous or discrete frequency representation. However, the primary bottleneck of Ly$\alpha$ radiative transfer is resonant scattering within the core of the line profile. Thus, frequency diffusion out of the core is usually necessary for spatial diffusion to occur, and applying Fick's law without the Fokker-Planck approximation provides only a marginal performance improvement. Still, a discrete frequency grouping MC implementation can be significantly faster than an equivalent continuous frequency version because redistribution across bins emulates the common core-skipping technique. Furthermore, even with core-skipping the discrete implementation takes advantage of precomputed redistribution matrix elements instead of on-the-fly computations. Therefore, we briefly comment on the discrete frequency grouping method. We define the redistribution function averaged over incoming and outgoing bins by
\begin{equation} \label{eq:R_discrete}
  R_{j' \rightarrow j} = \frac{1}{\Delta x_j \Delta x_{j'}} \iint_{j\,j'} R_{x' \rightarrow x}\, \text{d}x \text{d}x' \, .
\end{equation}
When a scattering event occurs the new frequency bin is efficiently drawn from this discrete probability distribution function. Unlike our Fokker-Planck DDMC scheme, frequency grouping is completely non-local and robust over all frequencies as long as $\Delta x_j \lesssim 1$, corresponding to a resolution requirement for redistribution in the wings. The advantage of the MC approach over similar discrete methods is that the redistribution matrix does not need to be inverted. However, the matrix elements themselves can be quite expensive to calculate, even with high-performance numerical libraries. To allow for variation of runtime parameters such as bin sizes and temperature, we can utilize the approximation from \citet{Gouttebroze_1986} to obtain $R_{x' \rightarrow x}$. We can then implement a two-dimensional adaptive quad tree integration scheme to obtain the final discretized version. This allows minimal overhead for the evaluation of equation~(\ref{eq:R_discrete}) and a significant speedup over continuous frequency MCRT.

However, due to the resolution requirement this approach becomes increasingly costly at high opacities as more photon packets are needed for convergence. Furthermore, the pre-computation presents a computational challenge for temperature- and angle-dependent redistribution in realistic simulation environments. Fortunately, the elements only need to be calculated once, although an efficient data representation may be necessary. For example, a uniform tabulation could have a large memory overhead, e.g. resolving $\{\Delta x,\Delta x'\} = 0.1$, $\Delta \log T = 0.1$, and $\Delta \cos \theta = 0.01$ requires $\sim 10^{11}$ elements, or $\sim$TB of data. In any case, initial tests for isothermal slabs with isotropic scattering are quite promising. Therefore, the main advantage of our new DDMC method for resonant lines is the computational efficiency, no costly redistribution matrix calculations, and convenient interfacing within existing MCRT codes. Still, depending on the problem it can be convenient to hybridize any combination of the three extensions discussed: our DDMC scheme, a continuous frequency scheme, and a discretized frequency redistribution scheme.

\subsection{Speedup for 3D simulations}
\label{sec:speedup}
In Section~\ref{sec:performance}, we demonstrated the improved performance of the rDDMC method over standard MCRT in highly idealized setups. To check whether similar speedups can be obtained in more realistic environments where regions of high and low optical depth exist, we have implemented the hybrid scheme for \textit{spatial} transport described in Section~\ref{sec:hybrid}, while maintaining a pure rDDMC approach for frequency redistribution. We note that this makes it difficult to reproduce the characteristic asymmetry of the red/blue Ly$\alpha$ peak heights induced by outflowing/inflowing gas. This is because the bin containing the core of the line acts as a barrier to leakage across the core. Therefore, it will be necessary to implement a robust hybrid \textit{frequency} scheme as well, most likely with adaptive kernel resolution in the wings and boundary conditions for conversion to MCRT in the core, as well as when at high temperatures the Doppler width exceeds the desired resolution. For our proof of concept implementation, we employ a gridless tophat kernel for exact Doppler-shifting of the bin edges when moving across cell interfaces. We also maintain a uniform resolution in real frequency characterized by the velocity offset from line centre, $\Delta v = c \Delta \lambda / \lambda_0$. We expect to improve our rDDMC implementation as needed for more realistic models of galaxies, black holes, the IGM, or stellar and planetary atmospheres.

\begin{figure}
  \centering
  \includegraphics[width=\columnwidth]{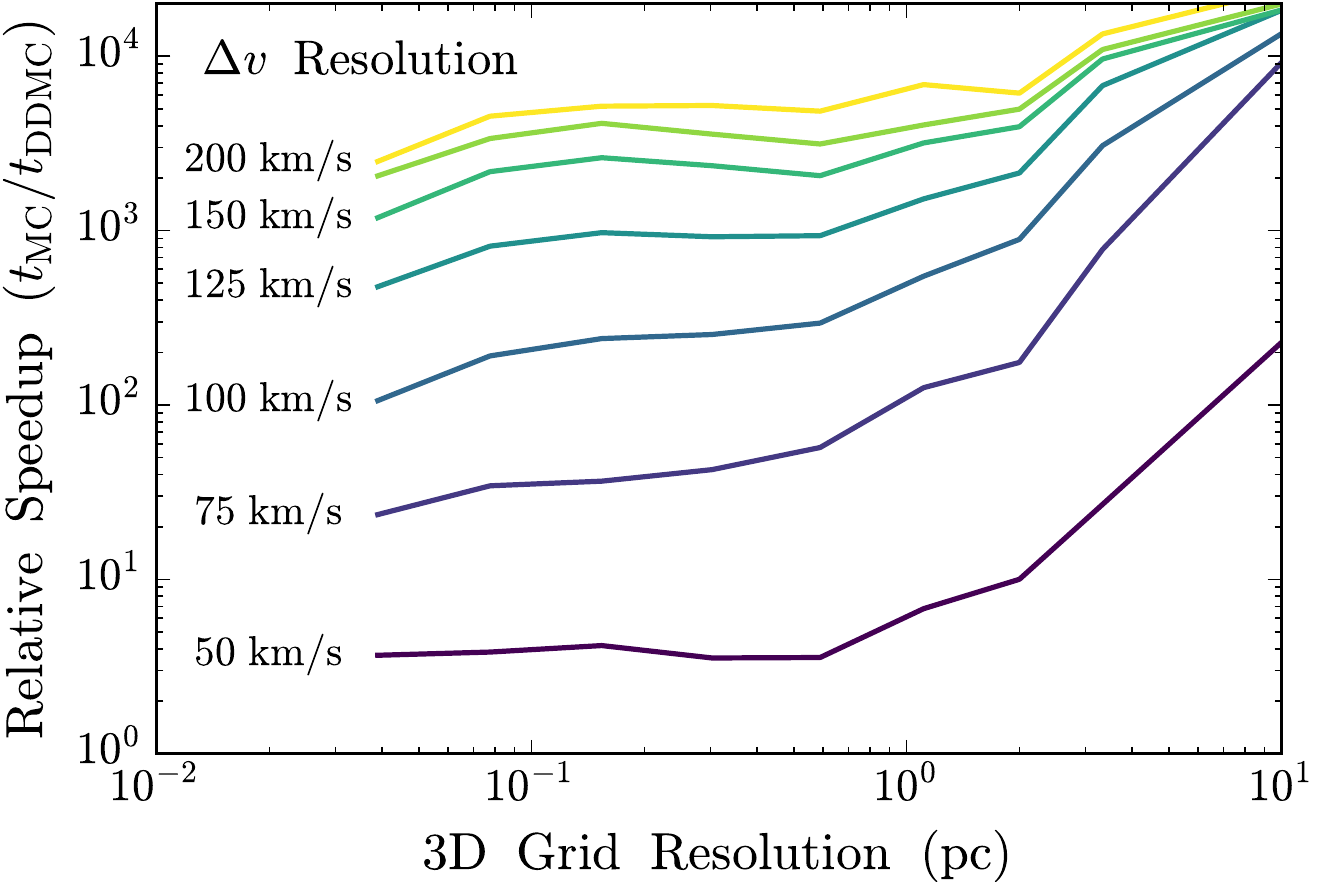}
  \caption{Relative speedup obtained with the rDDMC method compared to MCRT with continuous frequency redistribution ($t_\text{MC}/t_\text{DDMC}$). For details about the 3D cosmological atomic cooling halo setup see Section~\ref{sec:speedup} and \citet{Smith_DCBH_2017}. The rDDMC implementation employs hybrid \textit{spatial} transport (see Section~\ref{sec:hybrid}), while the MCRT utilizes aggressive nonlocal core-skipping \citep[\colt;][]{Smith_2015}. The simulation comparison is performed for Cartesian grids with varying spatial resolution from $\approx 0.04$--$10$\,pc and frequency resolution from $50$--$200\,\text{km\,s}^{-1}$.}
  \label{fig:timings}
\end{figure}

We apply our implementation to an \textit{ab initio} cosmological zoom-in simulation resolving the collapse of a primordial gas cloud in an atomically cooling dark matter halo. The simulation models the assembly of a direct collapse black hole (DCBH), a system expected to require Ly$\alpha$ radiation hydrodynamical studies. We refer the reader to \citet{Smith_DCBH_2017} for specific details and Ly$\alpha$ MCRT analysis related to the simulation data taken from \citet{Becerra_2018a,Becerra_2018b}. The original data is represented by a Voronoi tessellation of particles. However, for our comparison we deposit the central 10\,pc region onto Cartesian grids of varying spatial resolution up to $n_{\{x,y,z\}} = 2^8 + 1 = 257$, where as before we enforce an odd number of bins to ensure the symmetry of inserting photons at the origin. The highest resolution grid has a density dynamic range of almost five orders of magnitude. We then perform post-processing MCRT and rDDMC on the same initial conditions, finding that both methods converge to the highest resolution result. Following \citet{Smith_DCBH_2017}, we adopt the model with a neutral hydrogen fraction of $x_\text{\HI} = n_\text{\HI}/n_\text{H} \approx 10^{-4}$, which results in an angular-averaged line flux with a separation between the blue and red peaks of approximately $600\,\text{km\,s}^{-1}$, i.e. the emergent spectra are similar to the curve labeled `Central Source' in figure~10 of \citet{Smith_DCBH_2017}. The highest temperature is around $T \sim 10^4$\,K so the frequency discretization should be greater than $2\,v_\text{th} \approx 25\,\text{km\,s}^{-1}$ to maintain accuracy in the Fokker-Planck approximation. Therefore, we run rDDMC simulations with frequency resolution from $\Delta(\Delta v) = 50$--$200\,\text{km\,s}^{-1}$. The resulting speedups compared to a full MCRT implementation with aggressive nonlocal core-skipping \citep[\colt;][]{Smith_2015}, are shown in Fig.~\ref{fig:timings}. It is apparent that rDDMC can significantly outperform MCRT, although this likely depends on the specific application and code implementation. Still, this initial example is encouraging and we will continue to develop the rDDMC scheme for future applications on adaptive grids and unstructured mesh geometries.

We note that the quoted speedups apply to solving the radiative transfer equation for the local radiation field. However, it is often necessary to construct surface brightness images along particular lines of sight, which is typically done in one of two ways. Firstly, directional binning of escaping photons, which converges slowly as only a small fraction of photons fall into the differential solid angles required for high-resolution images. Still, the rDDMC speedup automatically applies here. Secondly, the next-event estimation or peel-off method uses weighted ray-tracing from every scattering event or a subset thereof to construct high signal-to-noise images of the Ly$\alpha$ photosphere. A hybrid rDDMC implementation transitioning to MCRT in lower opacity regions would produce essentially the same result as spatially discrete photon packets immediately contribute less than one per cent of their original weight, i.e. $\sim e^{-\tau_{\nu,\text{cell}}}$. However, for consistency we suggest sampling a random position within the cell and perform ray-tracing with early termination, i.e. discarding photons with negligible weight for computational efficiency. However, this method should be carefully tested as we anticipate the need for correction factors to account for the reduced number of physical scattering events. Finally, depending on the application it may be possible to embed information within the grid itself, e.g. $E(\bmath{r},\nu)$, to reconstruct observables in an additional post-processing step.

\section{Conclusions}
\label{sec:conc}
Careful analysis of radiative transfer in the Fokker-Planck approximation has provided us with additional theoretical and practical insight into Ly$\alpha$ transport. We have presented several analytical approximate solutions of the transfer equation. For further progress in more general environments we turn to simulations. Although there are many computational methods for line scattering problems, MCRT has emerged as the most common tool due to its susceptibility to simply and robustly incorporate additional physics (e.g., absorption by dust). Here, we have developed a novel extension of the DDMC method for resonance lines. Although we have focused on Ly$\alpha$ radiative transfer, the methodology is applicable to other lines in optically thick environments. The central idea is to skip scatterings in which either the mean free path is smaller than the cell size ($k_{i,j} \Delta z_i \gg 1$), or the typical frequency redistribution is similarly unresolved ($\Delta x_j \gg 1$).

We have shown that the performance characteristics under simple setups are straightforwardly interpretable. In particular, the DDMC runtime scales with the spatial and frequency resolution rather than the number of scatterings as in traditional MCRT. Therefore, our method successfully breaks the high opacity efficiency barrier inherent in MCRT and the DDMC framework can be readily incorporated into existing Ly$\alpha$ MCRT codes. 

The immediate frontiers include the full 3D Ly$\alpha$ RHD coupling in extremely high optical depth environments. However, even in cases where Ly$\alpha$ radiation pressure is not dynamically important, observational signatures may depend on the highly time-variable hydrodynamical coupling, which governs the escape of Ly$\alpha$ radiation from galaxies at intermediate and very high redshift \citep[e.g.][]{Finkelstein_2016}. Self-consistently incorporating Ly$\alpha$ and other forms of radiation into simulations with dusty, turbulent flows will provide a more complete understanding of these situations. 

In future work, we will apply our DDMC method to more complex 3D environments. Additionally, the DDMC method may be employed to explore other fundamental aspects of Ly$\alpha$ radiative transfer, such as dust extinction, recoil, collisional (de)excitation, cosmological expansion, turbulence, and other effects. Computational advances will thus assist in the interpretation of the high-precision data from next-generation telescopes.

\section*{Acknowledgements}
We thank the referee Peter Laursen for constructive comments that improved the content of this paper. This material is based upon work supported by a National Science Foundation Graduate Research Fellowship for AS. We acknowledge support from NSF grant AST-1413501.



\bibliographystyle{mnras}
\bibliography{bibliography/biblio}


\appendix

\onecolumn
\section{Approximations of the redistribution function}
\label{appendix:redistribution}
We now provide specific forms for the Ly$\alpha$ redistribution function. In particular, under isotropic coherent scattering without recoil, the redistribution function for the Voigt profile is \citep{Unno_1952,Hummer_1962}
\begin{equation}
  R_\text{II-A}(x,x') = \frac{1}{\pi H(x',a)} \int_\zeta^\infty e^{-u^2} \left[ \tan^{-1} \left(\frac{\underline{x} + u}{a}\right) - \tan^{-1} \left(\frac{\bar{x} - u}{a}\right) \right]\,\text{d}u \, ,
\end{equation}
where $\bar{x} = \max(x, x')$, $\underline{x} = \min(x, x')$, and $\zeta = (\bar{x} - \underline{x}) / 2 = |x - x'| / 2$. The limit as $a \rightarrow 0$ corresponds to the case with only Doppler broadening, so we may approximate the redistribution function in the core by
\begin{equation}
  R_\text{II-A,core}(x,x') \approx R_\text{I-A}(x,x') = \frac{\sqrt{\pi}}{2} e^{{x'}^2} \text{erfc}|\bar{x}| \, , \end{equation}
where $|\bar{x}| = \max(|x|,|x'|)$. In this approximation core photons have an equal probability of being scattered in the range $|x| < |x'|$ with an exponentially decreasing probability of being scattered to other frequencies. On the other hand, in the limit of large frequencies $|x+x'| \rightarrow \infty$ the approximate redistribution function in the wing is
\begin{equation}
  R_\text{II-A,wing}(x,x') \approx \left( \frac{2x'}{x+x'} \right)^2 \left[ \frac{e^{-\zeta^2}}{\sqrt{\pi}} - \zeta\,\text{erfc}(\zeta) \right] \, .
\end{equation}
The angular-averaged redistribution function under dipole scattering for the Voigt profile is given by
\begin{equation}
  R_\text{II-B}(x,x') = \frac{3 a}{8 \pi H(x',a)} \int_\zeta^\infty e^{-u^2} \int_{\bar{x} - u}^{\underline{x} + u} \left[ 3 - \left(\frac{x-t}{u}\right)^2 - \left(\frac{x'-t}{u}\right)^2 + 3 \left(\frac{x-t}{u}\right)^2 \left(\frac{x'-t}{u}\right)^2 \right] \frac{\text{d}t\,\text{d}u}{t^2 + a^2} \, .
\end{equation}
The limit as $a \rightarrow 0$ provides the corresponding approximation in the core, equivalent to $R_\text{I-B}(x,x')$,
\begin{equation}
  R_\text{II-B,core}(x,x') \approx \frac{3}{8} e^{{x'}^2} \left[ \frac{\sqrt{\pi}}{2} \text{erfc}|\bar{x}| \left( 3 + 2 x^2 + 2 {x'}^2 + 4 x^2 {x'}^2 \right) - e^{-|\bar{x}|^2} |\bar{x}| \left( 2 |\underline{x}|^2 + 1 \right) \right] \, ,
\end{equation}
where $|\underline{x}| = \min(|x|,|x'|)$. This modifies the redistribution because photons are preferentially forward- and back-scattered in the core. Finally, we calculate the limiting behavior of the redistribution function in the wing as
\begin{equation}
  R_\text{II-B,wing}(x,x') \approx \left( \frac{2x'}{x+x'} \right)^2 \left[ \left( \frac{11}{10} + \frac{8}{5} \zeta^2 + \frac{4}{5} \zeta^4 \right) \frac{e^{-\zeta^2}}{\sqrt{\pi}} - \left( \frac{3}{2} + 2 \zeta^2 + \frac{4}{5} \zeta^4 \right) \zeta\,\text{erfc}(\zeta) \right] \, .
\end{equation}
Additional references on the theoretical properties and practical computation of the redistribution function include: \citet{Ayres_1985}, \citet{Gouttebroze_1986}, \citet{Uitenbroek_1989}, \citet{Leenaarts_2012}, and \citet{Harutyunian_2016}. Furthermore, the effect of recoil on the redistribution function is considered by \citet{Field_1959} and \citet{Basko_1981}.


\bsp	
\label{lastpage}
\end{document}